# Enhancing Carrier Mobility by Remote Phonons

Chenmu Zhang*, Yuanyue Liu*

*Texas Materials Institute and Department of Mechanical Engineering,*

*The University of Texas at Austin, Austin, Texas 78712, USA*

*zcmben2014@gmail.com, Yuanyue.liu@austin.utexas.edu*

**Abstract:**

Remote phonons from dielectrics are generally believed to degrade carrier mobility in adjacent semiconductors through Fröhlich scattering of polar-optical phonons (POPs). Here, we show that remote phonons can instead enhance the mobility in van der Waals (vdW) heterostructures. We developed a first-principles computational framework to evaluate these remote phonon effects. Applying our approach to monolayer InSe semiconductor encapsulated by h-BN dielectric layers, we show that electron mobility is enhanced due to coupling between POPs in InSe and h-BN, which gives rise to a new collective phonon mode where the dielectric Fröhlich potential partially cancels that in the semiconductor. Guided by this mechanism, we further identified additional dielectrics that yield similar mobility enhancement. This work offers not only an effective computational method to evaluate the remote phonon effects but also insights into mobility engineering for next-generation electronics.

In electronics, semiconductors are often surrounded by dielectrics, which host "remote" phonons that can substantially affect carrier transport in adjacent semiconductors [1]. For instance, in metal-oxide-semiconductor field-effect transistors, carrier mobility of the Si inversion layer is significantly reduced by remote phonons from high-$\kappa$ dielectrics, such as $HfO_2$ and $La_2O_3$ [2, 3]. This mobility reduction is primarily caused by polar optical phonons (POPs), which scatter electrons in semiconductors via long-range Coulomb interactions (Fröhlich scattering). Remote phonon effects become even more pronounced in two-dimensional (2D) materials due to their atomic thinness and high susceptibility to the environment. Although experimentally isolating and quantifying remote phonon effects is challenging due to complications from other effects, such as scattering from remote defects, it is generally believed that remote phonons degrade carrier mobility. [1, 4-6],

Theoretically, quantitatively predicting how the semiconductor transport is affected by the surrounding dielectric is challenging due to several entangled contributions, as illustrated in Figure 1a: (1) Dielectric screening (DS) of the Fröhlich potential of the semiconductor POPs; (2) Remote POP scattering (RPS), where dielectric POPs introduce additional scattering to the electrons in the semiconductor; (3) Interlayer POP coupling (IPC), which hybridizes the phonons across the heterostructure and forms new collective POP modes. The latter two are considered as remote phonon effects. Prior works usually use simplified models that miss one or more contributions [6], neglect the anisotropic dielectric response typical of 2D materials [7,8], or assume the dielectric is continuous with limited atomistic information[4,9]. Although in principle these limitations can be overcome by direct first-principles calculations of the full heterostructure using methods such as density functional perturbation theory (DFPT), such calculations are often computationally prohibitive due to (1) lattice mismatch, which requires a large number of atoms, and (2) high computational cost of electron-phonon interaction. Moreover, such calculations include all the effects of dielectrics together, and it is unclear how to separate them for mechanistic understanding.

Here, we present a computational framework to quantify the effects of dielectrics on the carrier transport in the semiconductor. This method is accurate (validated against DFPT), and efficient (can calculate a large heterostructure). Using this method to study monolayer InSe semiconductor encapsulated by h-BN dielectric layers, we find that the remote phonons instead enhance the carrier mobility, in contrast to the conventional belief. This is because of the coupling between POPs in InSe and h-BN, which gives rise to a new collective phonon mode where the h-BN Fröhlich potential partially cancels that in InSe. Guided by this mechanism, we screen the 2D materials database and discover additional dielectrics that yield similar mobility enhancement. Therefore, our work offers an effective method to evaluate the dielectrics effects as well as insights into mobility engineering for next-generation electronics.

Our method uses DFPT to obtain the dielectric and dynamical properties for each individual layer in the heterostructure, and then integrate them to obtain the properties of the full system. Specifically, we leverage the Quantum Electrostatic Heterostructure (QEH) model [10,11] and separate the dynamical matrix $D(\mathbf{q})$ for the heterostructure as:

$$D(\mathbf{q}) = D^{\mathrm{a}}(\mathbf{q}) + D^{\mathrm{na}}(\mathbf{q}) \qquad (1)$$

where $D^{\mathrm{a}}$ and $D^{\mathrm{na}}$ are analytic and nonanalytic contributions [12,13], corresponding to short-range and long-range interactions, respectively. By assuming that interlayer coupling is only mediated by long-range Fröhlich potential induced by POPs, $D^{\mathrm{a}}$ is a diagonal block matrix with each block diagonalizable independently per layer. Diagonalizing $D^{\mathrm{a}}(\mathbf{q})$ for each layer yields layer-dependent frequencies $\omega_{\mathbf{q}l\nu}$ and eigenvectors $e_{\mathbf{q}l\nu}$, where $l\nu$ denotes phonon mode $\nu$ within layer $l$. By taking $e_{\mathbf{q}l\nu}$ as the basis set and approximating $\omega_{\mathbf{q}l\nu}$ by its values at zone-center ($\mathbf{q} = 0$), the full dynamical matrix simplifies to:

$$D_{l\nu,l'\nu'}(\mathbf{q}) = \delta_{ll'}\delta_{\nu\nu'}\omega_{0l\nu}^{2} + e^{2} z_{l\nu} z_{l'\nu'}^{*} \frac{|\mathbf{q}|^{2}\,\overline{W}_{ll'}(\mathbf{q})}{\sqrt{A_l A_{l'}}} \qquad (2)$$

where $\omega_{0l\nu}$ is the zone-center phonon frequency calculated from the density functional perturbation theory [12,13], $A_l$ is the unit cell area of layer $l$, and $\overline{W}_{ll'}$ is the screened Coulomb interaction between layers $l$ and $l'$ evaluated by the QEH model. $z_{l\nu}$ is the effective Born effective charge from collective atom displacements $e_{\mathbf{q}l\nu}$:

$$z_{l\nu} = \sum_{\alpha i} \frac{(\hat{\mathbf{q}} \cdot \mathbf{Z}_{l\alpha})_i e_{\mathbf{q}l\nu;\alpha i}}{\sqrt{M_{l\alpha}}} \quad (3)$$

where $\hat{\mathbf{q}}$ is the unit vector along the phonon wavevector, $i$ is the cartesian index, $M_{l\alpha}$ is the mass of atom $\alpha$ in the unit cell within layer $l$, and $\mathbf{Z}_{l\alpha}$ is its Born effective charge tensor.

By diagonalizing the dynamical matrix $D$ again for the entire heterostructure (Eq. 2), we obtain a new set of frequencies $\tilde{\omega}_{\mathbf{q}\mu}$ and eigenvectors $\tilde{e}_{\mathbf{q}\mu}$ for the collective modes:

$$\sum_{l'\nu'} D_{l\nu,l'\nu'}(\mathbf{q}) \tilde{e}_{\mathbf{q}\mu;l'\nu'} = \tilde{\omega}_{\mathbf{q}\mu}^2 e_{\mathbf{q}\mu;l\nu} \quad (4)$$

where the new index $\mu$ labels collective modes, with $\tilde{e}_{\mathbf{q}\mu;l\nu}$ quantifying contributions from uncoupled mode $\nu$ from individual layer $l$. With $\tilde{\omega}_{\mathbf{q}\mu}$ and $\tilde{e}_{\mathbf{q}\mu}$, the Fröhlich electron-phonon coupling (EPC) matrix element is calculated as:

$$g_{l\mu}^{\mathrm{F}}(\mathbf{k},\mathbf{q}) = \sum_{l'\nu} \frac{-i}{\sqrt{A_l A_{l'}}} \frac{\tilde{e}_{\mathbf{q}\mu;l'\nu}}{\sqrt{2\tilde{\omega}_{\mathbf{q}\mu}}} |\mathbf{q}| W_{l'l}(\mathbf{q}) z_{l'\nu} \quad (5)$$

where $W_{l'l}$ is the screened potential in layer $l$ induced by charges from layer $l'$ (defined slightly different from $\bar{W}_{ll'}$ in Eq. 2; see Note S2 in the SM), as calculated using the QEH model.

Once $g^{\mathrm{F}}$ is obtained, we employ the established procedures [14, 15] to calculate the Fröhlich-limited relaxation time $\tau^{\mathrm{F}}$ and the corresponding Fröhlich mobility $\mu^{\mathrm{F}}$. Then total mobility $\mu$ is then computed using Matthiessen's rule: $1/\mu = 1/\mu^{\mathrm{F}} + 1/\mu^{\mathrm{NF}}$, where $\mu^{\mathrm{F}}$ and $\mu^{\mathrm{NF}}$ are Fröhlich and non-Fröhlich mobility, respectively. Assuming that the surrounding layers modify only $\mu^{\mathrm{F}}$, we can compute dielectric-dependent $\mu$ using the same $\mu^{\mathrm{F}}$ for each heterostructure. Detailed derivations of $g^{\mathrm{F}}$, dielectric response using the QEH model, and mobility calculations are provided in the Supporting Information (SI).

To validate our model, we performed direct DFPT calculations on a $ZrS_2$-$HfS_2$ bilayer, and compared the EPC matrix elements with those predicted from our model. This system was chosen due to their matched lattice constant (which allows for a small cell and DFPT calculation over multiple q points) and their availability in QEH dielectric database. As shown in Figure 2, our model agrees well with the DFPT results, regardless of the interlayer distance (relaxed with van der Waals interaction[16] or artificially increased). For comparison, we also included the EPW interpolated EPC results, which show a noticeable deviation from the DFPT results. This is because currently EPW interpolation relies on dielectric property modeling of isolated monolayers[17, 18], which fails in thick heterostructures where dielectric screening is spatially inhomogeneous. More details are provided in Note S4 in the SI.

Our framework has several key features: (1) By incorporating the $\overline{W}_{ll'}$ from the QEH model, it captures the nontrivial, anisotropic dielectric screening of realistic multilayer stacks. (2) Diagonalizing the full heterostructure dynamical matrix naturally captures frequency renormalization and eigenvector hybridization of POPs induced by interlayer electrostatics. (3) This model facilitates the decomposition of dielectrics effects on mobility into DS, RPS and IPC contributions, by selectively truncating the dynamical matrix. Specifically, by setting $D_{lv,l'v'} = \delta_{l0}\,\delta_{l'0}\,D_{lv,l'v'}$, we obtain the results with DS only (i.e. neglecting the remote phonon contributions). With $D_{lv,l'v'} = \delta_{ll}\,D_{lv,l'v'}$, we add the RPS (DS + RPS), and using the full dynamical matrix $D$ captures all effects (DS + RPS + IPC). These three models enable a systematic analysis of dielectrics effects on mobility in the following discussion.

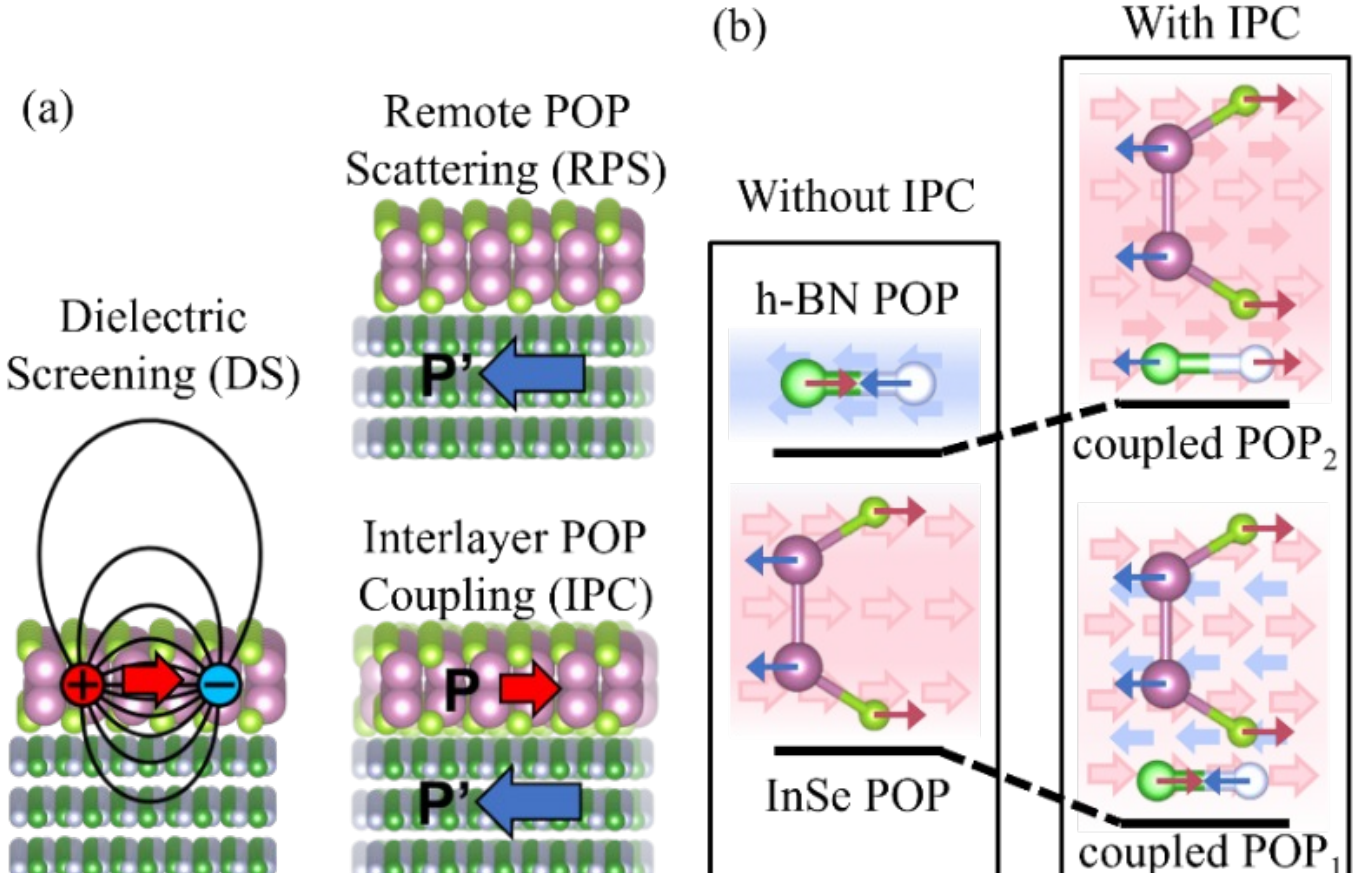

**Figure 1** (a) Schematic illustration of three effects of surrounding layers in a vdW heterostructure on carrier transport: dielectric screening (DS), remote POP scattering (RPS) and interlayer POP coupling (IPC). Arrows indicate POP vibrations and the resulting electrostatic fields. (b) Two-level model diagram for h-BN and InSe POPs with and without IPC. IPC hybridizes POPs for h-BN and InSe, forming new collective modes with shifted energy levels (black lines), different vibrational phases (arrows on atoms), and weakened or strengthened Fröhlich potential (arrows in the background).

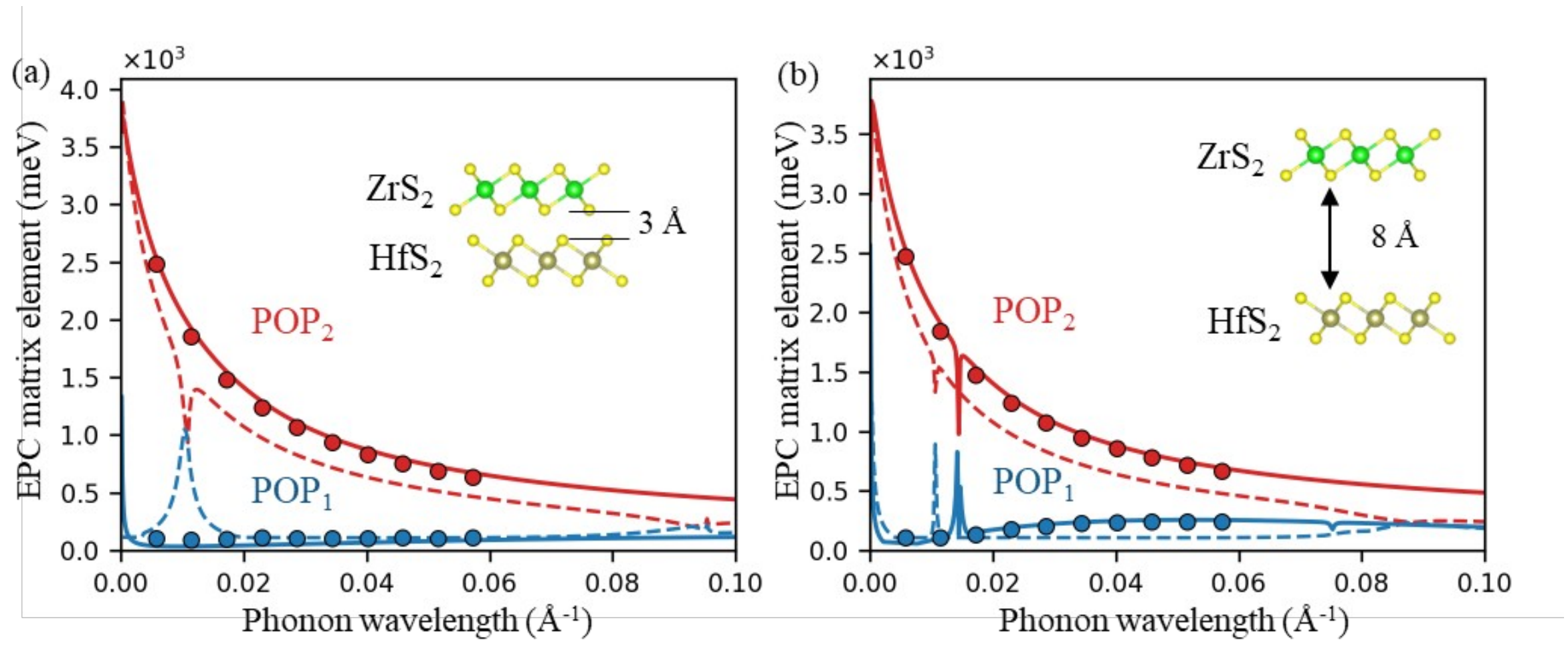

**Figure 2** Electron-phonon-coupling (EPC) matrix elements of $ZrS_2$ electrons in the $ZrS_2/HfS_2$ bilayer heterostructure, computed from DFPT (circles), EPW interpolation (dashed lines), and our model (solid lines). Red and blue curves correspond to the two hybrid polar optical phonon (POP) modes. (a) Bilayer with fully relaxed distance (b) Bilayer with expanded distance. The dips in the EPC curves are artifacts of dynamical matrix diagonalization: eigenvectors mix near frequency-degenerate q-points.

We note that this framework focuses on the long-range Fröhlich interaction of polar optical phonons as the dominant mechanism for remote phonon scattering. Substrate effects on non-polar phonon modes, including flexural (ZA) acoustic modes[19-21], out-of-plane optical (ZO) modes[22], and acoustic modes coupled via long-range dynamical quadrupoles[14, 23], are not included. We choose to focus on the Fröhlich interaction because the remote phonon scattering is induced by the POP in the dielectrics and mainly affects the POP in the semiconductors. In addition, we assume the electronic band structure of individual layers is preserved upon stacking. Therefore, this method is not applicable to systems where the electron structure significantly reconstructs such as Moiré systems with flat bands[24].

We start by studying monolayer InSe semiconductor encapsulated by h-BN dielectric layers. InSe is a common 2D semiconductor with relatively high intrinsic electron mobility of 117 $cm^2V^{-1}s^{-1}$, primarily limited by its Fröhlich scattering [14, 25], and h-BN is a widely used 2D dielectric. As shown in Figure 3a, the electron mobility in InSe (green line) increases with the number of h-BN layers, saturating at around 260 $cm^2V^{-1}s^{-1}$ with 10 layers (10L) h-BN on each side, which is 2.2 times of that in freestanding InSe. To understand the origin of this mobility enhancement, we use three models to incrementally include the effects of DS, RPS and IPC. With DS only (blue line), mobility increases monotonically with the number of dielectric layers, due to the enhanced dielectric screening of the InSe intrinsic Fröhlich scattering by h-BN. When the RPS is introduced (orange line), the mobility slightly decreases; while further inclusion of IPC (green line) increases mobility notably. Importantly, the IPC-induced mobility enhancement outweighs the reduction from RPS, resulting in a net mobility increase of 30 $cm^2V^{-1}s^{-1}$. This example contradicts the conventional expectation that remote phonons always degrade mobility, and demonstrates that IPC can lead to an overall increase of mobility.

The mobility enhancement by IPC can be explained using a simplified two-level model. As illustrated in Figure 1b, isolated InSe has a lower POP frequency (0.02 eV at zone center) compared to h-BN (0.17 eV). Once IPC is incorporated, these two POP states hybridize, resulting in two new hybrid modes ($POP_1$ and $POP_2$ in Fig. 1b). The lower-energy hybrid $POP_1$ is dominated by the InSe POP but contains a small out-of-phase component of the h-BN POP vibration. The opposite phase of POPs generates counteracting dipolar electric fields, thereby suppressing Fröhlich scattering. Conversely, for the higher-energy hybrid mode $POP_2$, the vibrations of InSe and h-BN POPs are in-phase, which potentially enhances Fröhlich scattering. However, its high frequency (> 0.17 eV) significantly reduces phonon occupancy at room temperature, leading to a minimal impact on overall mobility. Consequently, the net effect of IPC is to increase InSe mobility. To validate this explanation, we evaluated Fröhlich EPC strength for InSe encapsulated by 5L h-BN. As shown in Figure 3b, inclusion of IPC indeed reduces EPC strength of the lower-energy mode (lower panel) while increasing it for the higher-energy mode (upper panel).

Figure 3c further shows the contribution of each POP mode in the heterostructure to the Fröhlich scattering rates. In the DS model (left panel), only the InSe LO mode contributes, producing a single scattering peak at the InSe POP frequency (~0.02 eV). When RPS is introduced (right panel), the InSe peak remains unchanged, but additional peaks from h-BN LO modes appear near ~0.17 eV. These h-BN peaks share the same LO frequency (identical to the h-BN monolayer LO frequency) but differ in magnitude because each h-BN layer sits at a different distance from InSe, leading to different screened Coulomb coupling strengths. When IPC is further included (Figure 3d, right panel), both the InSe and h-BN peaks change. The InSe scattering rate decreases, which is a direct consequence of the Fröhlich potential cancellation by the out-of-phase h-BN POP vibration. Meanwhile, the h-BN peaks split and redistribute due to the long-range Coulomb coupling between h-BN layers: the originally degenerate h-BN LO modes hybridize into collective modes with distinct frequencies. The highest-frequency collective mode, representing in-phase vibrations of all h-BN layers, produces the largest peak. More details on IPC effects on frequency and scattering rate are in Note S5 in the SI.

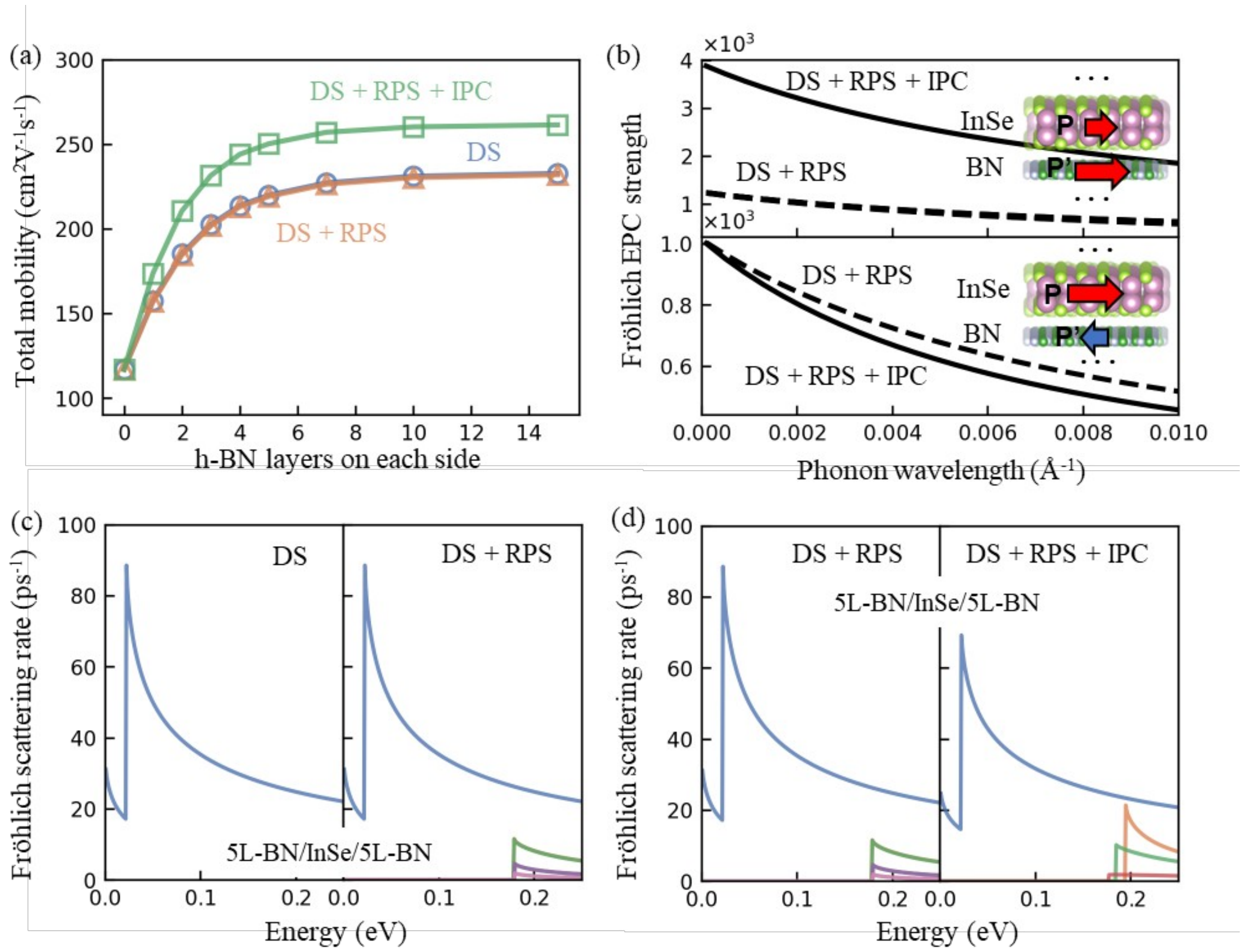

**Figure 3** (a) Total mobility of InSe encapsulated by h-BN layers, calculated using models with different contributions of dielectric screening (DS), remote POP scattering (RPS) and interlayer POP coupling (IPC) effects. See text for details of the calculation methods. (b) Fröhlich EPC strength of InSe encapsulated by 5 h-BN layers on each side, calculated with (solid lines) and without IPC (dashed lines). Insets illustrate the InSe and one h-BN layer in the heterostructure, with arrows indicating POP-induced electrostatic fields. (c-d) InSe electron Fröhlich scattering rates contributed by each POP mode in the 5L-BN/InSe/5L-BN heterostructure. Each curve represents a different POP mode.

The above studies focus on h-BN as dielectrics, which is commonly used but suffers from high leakage currents due to its relatively narrow bandgap and low permittivity [26]. Therefore, we explore other promising 2D dielectrics that might enhance the mobility. Using the Computational 2D Materials Database (C2DB) [27, 28], we screened over 4000 2D materials via a three-step process: (1) Select insulators with electronic bandgaps exceeding 4 eV, by using the more accurate HSE bandgap available in the database; (2) Exclude materials lacking "dynamical stability" and "stiffness stability" as labeled by the database; (3) Evaluate the lowest POP frequency at zone center ($\omega_0$) of 2D layers based on Born effective charge tensor and dynamical matrix provided by the database, and exclude materials with $\omega_0$ < InSe POP frequency (0.02 eV). The screening process eventually yields 37 potential 2D dielectrics, from which five (h-BN, $GeO_2$, $SnO_2$, $HfO_2$ and $ZrO_2$) are selected due to their availability of first-principles dielectric data in the QEH database (complete list of dielectrics can be found in Note S6 in the SI). The criterion $\omega_0$ > 0.02 eV ensures that IPC reduces Fröhlich scattering for the lower-energy

InSe mode, motivated by the insight provided by the two-level model above. In addition, another key insight from the two-level model is the strength of IPC: a large IPC strength leads to large POP hybridization and more suppression for Fröhlich scattering. Here we use $\Delta\omega$ to quantify IPC strength, which is defined as:

$$\Delta\omega = \frac{ez_{\mathrm{POP}}}{\sqrt{\alpha^{\parallel} A}} \tag{3}$$

where $e$ is the electron charge, $z_{\mathrm{POP}}$ is the mode-resolved Born effective charge corresponding to the POP vibration in dielectric (defined in Eq. 3), $\alpha^{\parallel}$ is the in-plane electronic polarizability of the 2D dielectric, and $A$ is its unit cell area. In the same unit of frequency, $\Delta\omega$ characterizes the LO-TO splitting of 2D materials [17, 18]. It is proportional to the off-diagonal coupling element in dynamical matrix and thus measures the IPC strength.

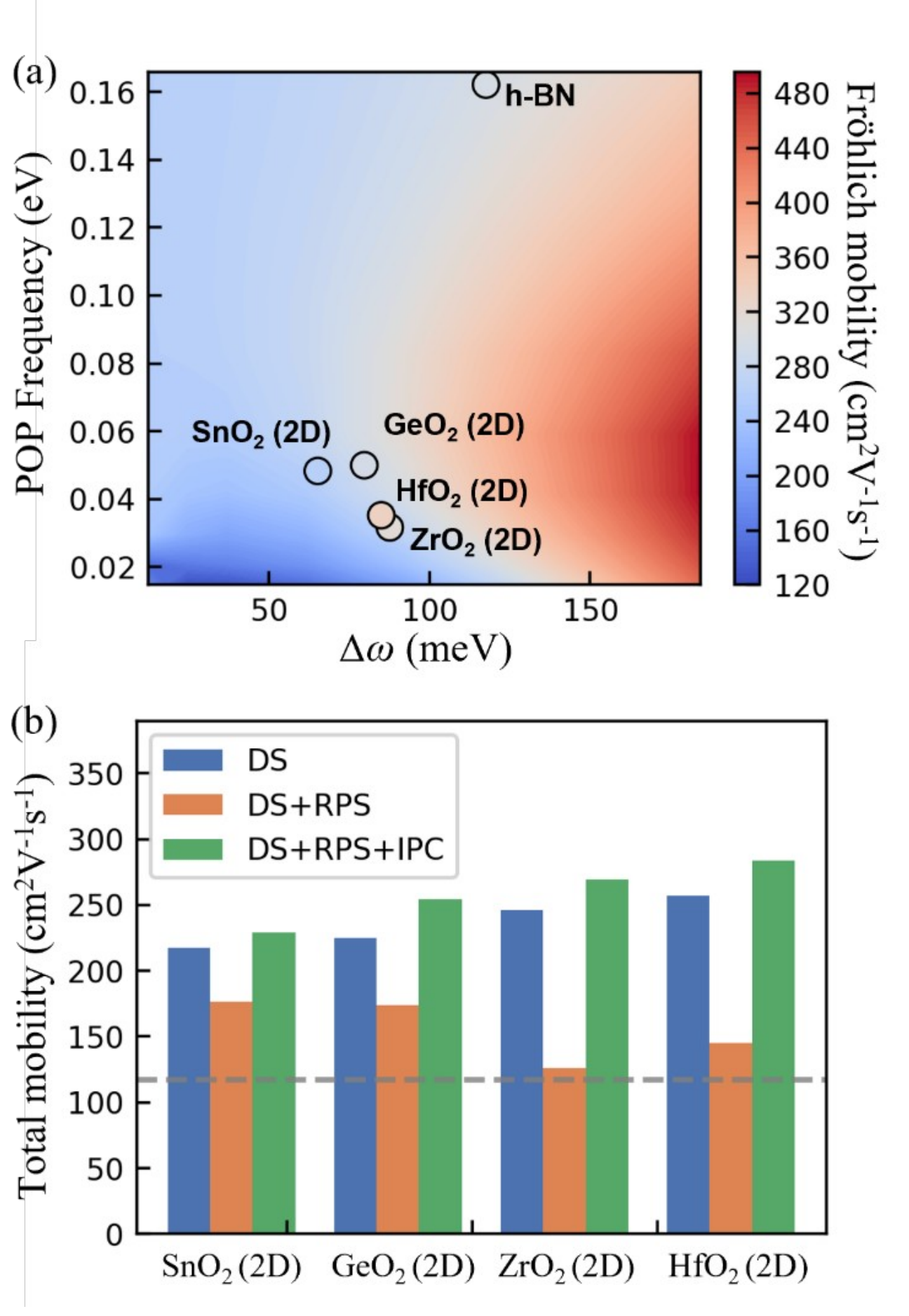

**Figure 4** (a) InSe Fröhlich mobility $\mu^{\mathrm{F}}$ with varied zone-center POP frequency $\omega_0$ and $\Delta\omega$ (see Eq. 6 for definition) of surrounding 2D dielectrics (7L on each side). Background color represents mobility calculated from a modified InSe/h-BN heterostructure model, with $\omega_0$ and $\Delta\omega$ of h-BN varied by scaling its dynamical matrices and Born effective charges, respectively. Selected promising 2D dielectric insulators with their $\omega_0$ and $\Delta\omega$ are indicated by dots, colored according to InSe $\mu^{\mathrm{F}}$ in corresponding heterostructures (7L each side). (b) Total mobility of InSe encapsulated by $SnO_2$, $GeO_2$, $ZrO_2$ and $HfO_2$ (7L each side), calculated using different models. Mobility of freestanding InSe monolayer (117 $cm^2V^{-1}s^{-1}$) is shown in grey for comparison.

Figure 4a displays the $\omega_0$ and $\Delta\omega$ of 2D h-BN, $SnO_2$, $GeO_2$, $ZrO_2$ and $HfO_2$, and Fröhlich mobility of monolayer InSe ($\mu^F$; indicated by colored dots) when encapsulated by these 2D dielectric layers (7L on each side). The background colors represent $\mu^F$ from a modified InSe/h-BN heterostructure model, where $\omega_0$ and $\Delta\omega$ of h-BN are continuously varied by scaling its dynamical matrices and Born effective charge tensors, respectively. Several key trends regarding the effects of $\omega_0$ and $\Delta\omega$ of dielectrics on $\mu^F$ are observable in Figure 4a: (1) For a fixed $\omega_0$, a larger $\Delta\omega$ typically yields a higher $\mu^F$. For instance, $GeO_2$ and $SnO_2$ have comparable $\omega_0$ (0.050 vs. 0.048 eV), but $GeO_2$ exhibits a larger $\Delta\omega$ (80 vs. 65 meV). This larger $\Delta\omega$ in $GeO_2$ increases its LO-TO splitting and thus the POP frequency near the zone center [17, 18], which reduces phonon occupancy for RPS. Simultaneously, it strengthens IPC and enhances Fröhlich potential cancellation; both effects suppress Fröhlich scattering and improve $\mu^F$; (2) Increasing $\omega_0$ at a fixed $\Delta\omega$ results in a non-monotonic variation of $\mu^F$. For example, while the $\omega_0$ values follow the order: $GeO_2$ (0.050 eV) > $HfO_2$ (0.035 eV) > $ZrO_2$ (0.031 eV), the corresponding InSe $\mu^F$ is $GeO_2$ (295 $cm^2V^{-1}s^{-1}$) < $ZrO_2$ (315 $cm^2V^{-1}s^{-1}$) < $HfO_2$ (335 $cm^2V^{-1}s^{-1}$), where $HfO_2$ leads to highest $\mu^F$ with a moderate $\omega_0$. This is due to that a very low $\omega_0$ increases phonon occupancy for RPS, which suppresses $\mu^F$; while a very high $\omega_0$ leads to a large POP energy mismatch, reducing the POP hybridization and the cancellation of Fröhlich potential according to the two-level model, which also degrades $\mu^F$. Consequently, achieving optimal $\mu^F$ in encapsulated InSe requires dielectrics with moderate $\omega_0$ and large $\Delta\omega$, which is more visible in the colormap of Figure 4a.

Figure 4b compares the total mobilities for InSe encapsulated by $SnO_2$, $GeO_2$, $ZrO_2$ and $HfO_2$ 2D layers using different models. From this comparison, we can quantify the individual contributions of DS, RPS and IPC. For all four dielectrics, DS significantly increases mobility from the isolated InSe value (117 $cm^2V^{-1}s^{-1}$; grey dashed line) to 217, 224, 245 and 256 $cm^2V^{-1}s^{-1}$, respectively, for $SnO_2$, $GeO_2$, $ZrO_2$ and $HfO_2$ encapsulation. The introduction of RPS notably reduces mobility, leading to 20%, 22%, 48% and 44% mobility reduction, respectively, compared to the DS-only model. Unlike h-BN, these dielectrics have lower $\omega_0$ and thus exhibit more significant RPS-induced mobility degradation. However, the lower $\omega_0$ also enhances POP hybridization due to the smaller frequency mismatch, which completely counteracts the RPS-induced degradation. This leads to final mobilities of 229, 254, 269 and 284 $cm^2V^{-1}s^{-1}$, for $SnO_2$, $GeO_2$, $ZrO_2$ and $HfO_2$ encapsulations, respectively, demonstrating the critical role of IPC in evaluating environmental effects on semiconductor mobility. We further evaluate the mobility enhancement for the complete list of 37 2D dielectric candidates from the C2DB. For this screening, we utilize the dielectric properties of h-BN to model the environmental screening, as full dielectric data for all candidates are not available in the QEH database. The highest predicted InSe mobility is 315 $cm^2V^{-1}s^{-1}$ which is 2.7 times of freestanding InSe. It should be noted that IPC-induced mobility enhancement also depends on the semiconductor properties. By substituting InSe with monolayer $MoS_2$, dielectric encapsulation decreases $MoS_2$ mobility by 6%-18%. This degradation is attributed to the lower $\omega_0$ and $\Delta\omega$ in $MoS_2$, which result in less effective phonon hybridization and potential cancellation. Detailed calculation results for the $MoS_2$ system are provided in Note S7 in the SI.

In this work, we develop an accurate and efficient first-principles method to evaluate the effects of the dielectrics on the carrier transport in the semiconductors in the vdW heterostructures. Our method explicitly incorporates the effects of dielectric screening, remote phonon scattering and interlayer phonon coupling, and also allows for separate evaluation of each effect. Using InSe encapsulated by h-BN layers as an example, our calculations reveal that remote phonons can enhance the electron mobility rather than degrading as conventionally thought. This enhancement originates from interlayer phonon coupling, which forms collective modes with reduced Fröhlich potentials that outweigh the remote phonon scattering. By screening a 2D materials database (C2DB), we identify additional dielectric candidates that yield similar mobility enhancement. The method and insights presented in this paper help advance the understanding and design of next-generation electronic and optoelectronic devices.

## Supporting Information

Detailed derivations of Fröhlich EPC matrix element in our model; phonon dispersion and EPC validation for $ZrS_2$/$HfS_2$ bilayer; screening results for 37 dielectric candidates; mobility of h-BN capsulated $MoS_2$.

## Code Availability

The code used in this work is openly available at https://github.com/cz2014/qeh-frohlich.

## Acknowledgement

This work is supported by NSF (2425545), Office of Naval Research (N00014-23-1-2400), and Welch Foundation (F-1959). The calculations were performed on the clusters of ACCESS, TACC and NREL.

**TOC**

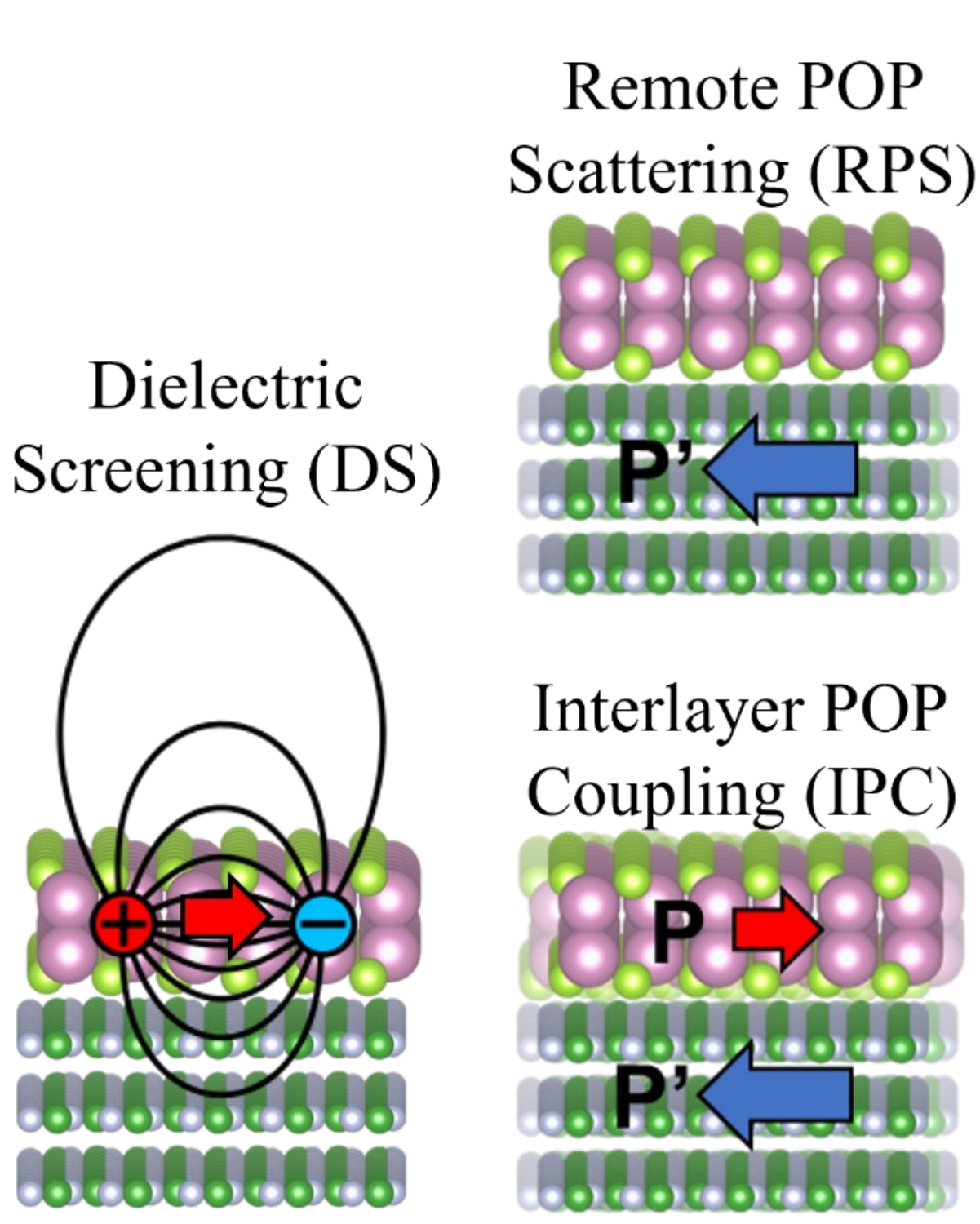

# 111Equation Chapter 1 Section 1Supporting Information:

## Enhancing Carrier Mobility by Remote Phonons

Chenmu Zhang*, Yuanyue Liu*

*Texas Materials Institute and Department of Mechanical Engineering,*

*The University of Texas at Austin, Austin, Texas 78712, USA*

*zcmben2014@gmail.com, Yuanyue.liu@austin.utexas.edu*

**Supplementary Note S1: Dynamical matrix and Fröhlich coupling in vdW heterostructure**

The Hamiltonian of lattice vibrations in a van der Waals (vdW) heterostructure can be written as:

$$H_{\text{tot}} = H^{a}_{\text{tot}} + V^{na}_{\text{tot}} = T_{\text{tot}} + V^{a}_{\text{tot}} + V^{na}_{\text{tot}} \tag{1}$$

where $T$ is the kinetic energy term, $V^a$ is the analytic potential term due to the short-range interaction and $V^{na}$ is the nonanalytic potential term arising from long-range Coulomb interactions [1, 2]. Here we assume that layers in heterostructure interact solely via vdW interactions so that the interlayer coupling in Eq. 1 only exist in the $V^{na}$ term. Therefore, the $H^a$ in Eq. 1 can be decomposed into contributions from individual layers:

$$H^{a}_{\text{tot}} = \sum_{l} H^{a}_{l} \tag{2}$$

and $H^a{}_l$ is:

$$H^{a}_{l} = \sum_{p\alpha i} -\frac{1}{2M_\alpha}\frac{\partial^2}{\partial h^2_{pl\alpha i}} + \frac{1}{2}\sum_{p\alpha i}\sum_{p'\alpha' i'} D^{a}_{l\alpha i,l\alpha' i'}(\mathbf{R}_p,\mathbf{R}_{p'}) h_{pl\alpha i} h_{p'l\alpha' i'} \tag{3}$$

where $p$ indexes the unit cells in the whole Born-von Kármán (BvK) supercell, $\alpha$ indexes the atoms within a unit cell, $i$ denotes the cartesian directions $x$, $y$, $z$; $h$ is the atomic displacement from their equilibrium positions, $\mathbf{R}$ is the unit cell coordinates, $M$ is the atomic mass and $D^a$ is the force constants due to short-range interactions. A standard solution to Eq. 3 can be obtained by transforming the coordinates $h_{pl\alpha i}$ into a set of complex normal coordinates $u_{\mathbf{q}l\nu}$:

$$\begin{aligned} u_{\mathbf{q}l\nu} &= \frac{1}{\sqrt{N_l}}\sum_{p\alpha i}\sqrt{M_{l\alpha}}\, h_{pl\alpha i}\, e^{-i\mathbf{q}\cdot\mathbf{R}_p} e_{-\mathbf{q}l\nu;\alpha i} \\ h_{pl\alpha i} &= \frac{1}{\sqrt{N_l}}\sum_{\mathbf{q}\in\Omega_l,\nu}\frac{1}{\sqrt{M_{l\alpha}}} u_{\mathbf{q}l\nu}\, e^{i\mathbf{q}\cdot\mathbf{R}_p} e_{\mathbf{q}l\nu;\alpha i} \end{aligned} \tag{4}$$

where $e_{\mathbf{q}l\nu}$ is the eigenvector of the analytic dynamical matrix $D^a(\mathbf{q})$:

$$\sum_{\alpha' i'} D^{a}_{l\alpha i,l\alpha' i'}(\mathbf{q})\, e_{\mathbf{q}l\nu;\alpha' i'} = \omega^2_{\mathbf{q}\nu} e_{\mathbf{q}l\nu;\alpha i} \tag{5}$$

and $D^a(\mathbf{q})$ is the Fourier transform of force constants:

$$D^a_{l\alpha i,l\alpha' i'}(\mathbf{q}) = \sum_p \frac{1}{\sqrt{M_\alpha M_{\alpha'}}} D^a_{l\alpha i,l\alpha' i'}(0,\mathbf{R}_p) e^{i\mathbf{q}\cdot\mathbf{R}_p} \tag{6}$$

Inserting $h$ from Eq. 4 into Eq. 3 and combining Eqs. 5 and 6, $H^a$ for the entire vdW heterostructure can be written as:

$$H^a_{\text{tot}} = \sum_l \sum_{\mathbf{q}\in Q_l,\nu} \frac{1}{2}\left(\frac{\partial}{\partial u_{\mathbf{q}l\nu}} \frac{\partial}{\partial u^*_{\mathbf{q}l\nu}} + \omega^2_{\mathbf{q}l\nu} u_{\mathbf{q}l\nu} u^*_{\mathbf{q}l\nu}\right) \tag{7}$$

Before explicitly expressing the interlayer coupling term $V^{na}$, here we provide essential details on the boundary conditions used in the vdW heterostructure system. We adopt a supercell commensurate with each layer as the BvK supercell. In this supercell, each layer contains $N_l = N_{l1} \times N_{l2}$ unit cells. The total area of the supercell is denoted by $A_{\text{tot}}$ and the area of the unit cell in layer $l$ is $A_l$, so that $A_{\text{tot}} = N_l A_l = N_{l'} A_{l'}$. For simplicity, we assume that the corresponding lattice vectors in different layers are mutually parallel (this is always possible by expanding the supercell). The relations of unit cell lattice vectors $\mathbf{a}_{li}$ and reciprocal lattice vectors $\mathbf{b}_{lj}$ between different layers are:

$$\begin{gathered} N_{li}\mathbf{a}_{li} = N_{l'i}\mathbf{a}_{l'i} \\ \frac{\mathbf{b}_{li}}{N_{li}} = \frac{\mathbf{b}_{l'i}}{N_{l'i}} \end{gathered} \tag{8}$$

The Bloch wave vectors $\mathbf{q}_l$ for each layer are on a uniform grid:

$$\mathbf{q}_l = \sum_j m_j \frac{\mathbf{b}_{lj}}{N_{lj}}, m_j = 0,\cdots, N_{lj}-1 \tag{9}$$

From Eq. 8, the $\mathbf{q}_l$ share common values around the zone center for all $l$ layers.

The Coulomb screened potential generated by the dipole induced by atomic displacement $h_{pl\alpha i}$ from layer $l$ is [3]:

$$V^c_l(\mathbf{r}; h_{pl\alpha i}) = \frac{-ie}{A_{\text{tot}}} \sum_{\mathbf{q}} W_l(\mathbf{q},z)(\mathbf{q}\cdot\mathbf{Z}^*_{l\alpha})_i h_{pl\alpha i} e^{i\mathbf{q}\cdot(\mathbf{r}-\mathbf{R}_p-\boldsymbol{\tau}_{l\alpha})} \tag{10}$$

Here, $W_l(\mathbf{q},z)$ is the screened potential with in-plane reciprocal lattice vector $\mathbf{q}$ and out-of-plane real space coordinate $z$, which can be calculated using the Quantum Electrostatic Heterostructure (QEH) model [4, 5] (see Note S2 for details). We drop the subscript $l$ on $\mathbf{q}$ since we are only interested in zone-center properties (with $W$ considered as a long-range interaction that vanishes for large $|\mathbf{q}|$), and the $\mathbf{q}_l$ have same values around the zone center. The interaction energy between the dipole from $h_{pl\alpha i}$ and the dipole from $h_{p'l'\alpha' i'}$ is:

$$V^c_{ll'}(h_{p'l'\alpha' i'}; h_{pl\alpha i}) = \frac{e^2}{A_{\text{tot}}} \sum_{\mathbf{q}} \frac{W_{ll'}(\mathbf{q}) + W_{l'l}(\mathbf{q})}{2} (\mathbf{q}\cdot\mathbf{Z}^*_{l'\alpha'})_{i'} (\mathbf{q}\cdot\mathbf{Z}^*_{l\alpha})_i h_{p'l'\alpha' i'} h_{pl\alpha i} e^{i\mathbf{q}\cdot(\mathbf{R}_{p'}-\mathbf{R}_p)} e^{i\mathbf{q}\cdot(\boldsymbol{\tau}_{l'\alpha'}-\boldsymbol{\tau}_{l\alpha})} \tag{11}$$

where $W_{ll'}(\mathbf{q})$ is the screened potential matrix obtained from the QEH model. Thus, $V^{na}$ in Eq. 1 can

be written as:

$$V_{\text{tot}}^{na} = \frac{1}{2}\sum_{l'p'\alpha'i'}\sum_{lp\alpha i} V_{ll'}^{c}(h_{p'l'\alpha'i'};h_{pl\alpha i}) \tag{12}$$

Inserting $h$ from Eq. 4 into Eq. 12 and applying orthogonal normalization relations:

$$\sum_{\mathbf{q}\in Q_l} e^{i\mathbf{q}\cdot\mathbf{R}_p} = N_l \delta_{p0}$$
$$\sum_{p\in P_l} e^{i\mathbf{q}\cdot\mathbf{R}_p} = N_l \delta_{\mathbf{q}0} \tag{13}$$

we can express $V^{na}$ in terms of the normal coordinates $u_{\mathbf{q}l\nu}$:

$$V_{\text{tot}}^{na} = \frac{e^2}{2}\sum_{\mathbf{q}}\sum_{ll',\nu\nu'} u_{\mathbf{q}l\nu}u_{\mathbf{q}l'\nu'}^{*} z_{l\nu} z_{l'\nu'}^{*} \frac{|\mathbf{q}|^2\,\overline{W}_{ll'}(\mathbf{q})}{\sqrt{A_l A_{l'}}} \tag{14}$$

where $\overline{W}_{ll'}(\mathbf{q}) = [W_{ll'}(\mathbf{q}) + W_{l'l}(\mathbf{q})]/2$ and $z_{l\nu}$ is given by:

$$z_{l\nu} = \sum_{\alpha i} \frac{(\hat{\mathbf{q}}\cdot\mathbf{Z}_{l\alpha}^{*})_i\, e_{\mathbf{q}l\nu;\alpha i}}{\sqrt{M_{l\alpha}}} \tag{15}$$

Writing Eq. 14 in terms of the dynamical matrix:

$$V_{\text{tot}}^{na} = \frac{1}{2}\sum_{\mathbf{q}}\sum_{ll'\nu\nu'} D_{l\nu,l'\nu'}^{na}(\mathbf{q})\, u_{\mathbf{q}l\nu}u_{\mathbf{q}l'\nu'}^{*} \tag{16}$$

Therefore, the dynamical matrix can be deduced by comparing Eqs. 14 and 16:

$$D_{l\nu,l'\nu'}^{na}(\mathbf{q}) = e^2 z_{l\nu} z_{l'\nu'}^{*} \frac{|\mathbf{q}|\,\overline{W}_{ll'}(\mathbf{q})}{\sqrt{A_l A_{l'}}} \tag{17}$$

With $H^a$ from Eq. 7 and $V^{na}$ from Eq. 16, now we have all the ingredients for the lattice vibration Hamiltonian of entire vdW heterostructure $H_{\text{tot}}$. Near the zone center, $H_{\text{tot}}$ can be expressed in terms of the normal coordinates $u_{\mathbf{q}l\nu}$:

$$H_{\text{tot}} = \frac{1}{2}\sum_{\mathbf{q}}\sum_{ll'\nu\nu'}\left[\delta_{ll'}\delta_{\nu\nu'}\frac{\partial}{\partial u_{\mathbf{q}l\nu}}\frac{\partial}{\partial u_{\mathbf{q}l\nu}^{*}} + \left(\delta_{ll'}\delta_{\nu\nu'}\omega_{\mathbf{q}l\nu}^2 + D_{l\nu,l'\nu'}^{na}(\mathbf{q})\right)u_{\mathbf{q}l\nu}u_{\mathbf{q}l\nu}^{*}\right] \tag{18}$$

which yields a new dynamical matrix for vdW lattice vibration that includes long-range interlayer phonon coupling:

$$D_{l\nu,l'\nu'}(\mathbf{q}) = \delta_{ll'}\delta_{\nu\nu'}\omega_{\mathbf{q}l\nu}^2 + e^2 z_{l\nu} z_{l'\nu'} \frac{|\mathbf{q}|^2\,\overline{W}_{ll'}(\mathbf{q})}{\sqrt{A_l A_{l'}}} \tag{19}$$

The frequency $\omega_{\mathbf{q}l\nu}$ is calculated by diagonalizing the analytic dynamic matrix $D^a$, which should be also analytic at the zone center. By approximate $\omega_{\mathbf{q}l\nu}$ using $\omega_{\mathbf{0}l\nu}$ which is calculated from $D^a(\mathbf{q}=0)$, we obtain the Eq. 2 in the main text.

Next, we diagonalize the $D$ in Eq. 19 to obtain the phonon eigenvectors and frequency of the vdW heterostructure. Similar to Eq. 5, the $\tilde{e}_{\mathbf{q}\mu;l\nu}$ and $\tilde{\omega}^2_{\mathbf{q}\mu}$ are the eigenvectors and eigenvalues of the dynamical matrix $D$, respectively:

$$\sum_{l'\nu'} D_{l\nu,l'\nu'}(\mathbf{q})\tilde{e}_{\mathbf{q}\mu;l'\nu'} = \tilde{\omega}^2_{\mathbf{q}\mu} e_{\mathbf{q}\mu;l\nu} \tag{20}$$

With $\tilde{e}_{\mathbf{q}\mu;l\nu}$, we are able to transform the normal coordinates $u_{\mathbf{q}l\nu}$ into a set of new complex normal coordinates $U_{\mathbf{q}\mu}$:

$$U_{\mathbf{q}\mu} = \sum_{l\nu} \tilde{e}^*_{\mathbf{q}\mu;l\nu} u_{\mathbf{q}l\nu} \tag{21}$$

The Hamiltonian of lattice vibration $H_{\text{tot}}$ is diagonal in the $U_{\mathbf{q}\mu}$ representation:

$$H_{\text{tot}} = \sum_{\mathbf{q}} \sum_{\mu} \frac{1}{2} \left( \frac{\partial}{\partial U_{\mathbf{q}\mu}} \frac{\partial}{\partial U^*_{\mathbf{q}\mu}} + \tilde{\omega}^2_{\mathbf{q}\mu} U_{\mathbf{q}\mu} U^*_{\mathbf{q}\mu} \right) \tag{22}$$

With the frequency of the hybrid phonon $\tilde{\omega}_{\mathbf{q}\mu}$, we can calculate the Fröhlich electron-phonon coupling (EPC) matrix element $\bar{g}^{\text{F}}$. Following the definition in Ref. [6], the $\bar{g}^{\text{F}}$ in the BvK boundary conditions is defined as:

$$\bar{g}^{\text{F}}_{l\mu}(\mathbf{k},\mathbf{q}) = \langle \psi_{l\mathbf{k}+\mathbf{q}} | \frac{-e}{\sqrt{2\tilde{\omega}_{\mathbf{q}\mu}}} \frac{\partial V^c_l(\mathbf{r})}{\partial U_{\mathbf{q}\mu}} | \psi_{l\mathbf{k}} \rangle \tag{23}$$

where $\psi_{l\mathbf{k}}$ is the wavefunction with wavevector $\mathbf{k}$ within layer $l$ and $V^c_l(\mathbf{r})$ is the total Fröhlich perturbation potential acting on layer $l$:

$$V^c_l(\mathbf{r}) = \sum_{p\alpha i} V^c_l(\mathbf{r}; h_{pl\alpha i}) \tag{24}$$

where $V^c_l(\mathbf{r};h_{pl\alpha i})$ is Fröhlich perturbation potential due to atomic displacement $h_{pl\alpha i}$ defined in Eq. 10. Inserting $h_{pl\alpha i}$ from Eq. 4 into Eq. 24 gives:

$$V^{\text{F}}_l(\mathbf{r}) = \frac{-ie}{A_l} \sum_{\mathbf{q}\nu} \frac{1}{\sqrt{N_l}} W_l(\mathbf{q},z) z_{l\nu} u_{\mathbf{q}l\nu} e^{i\mathbf{q}\cdot\mathbf{r}} \tag{25}$$

It immediately follows from Eq. 25 that:

$$\frac{\partial V^{\text{F}}_l(\mathbf{r})}{\partial u_{\mathbf{q}l\nu}} = \frac{-ie}{A_l} \frac{1}{\sqrt{N_l}} |\mathbf{q}| W_l(\mathbf{q},z) z_{l\nu} \tag{26}$$

Using Eq. 26 and Eq. 21, finally we can calculate $\bar{g}^{\text{F}}$ (defined in Eq. 23) by:

$$\bar{g}^{\mathrm{F}}_{l\mu}(\mathbf{k},\mathbf{q}) = \langle \psi_{l\mathbf{k}+\mathbf{q}} | \frac{-e}{\sqrt{2\tilde{\omega}_{\mathbf{q}\mu}}} \frac{\partial V_l^c(\mathbf{r})}{\partial U_{\mathbf{q}\mu}} | \psi_{l\mathbf{k}} \rangle$$

$$= \langle \psi_{l\mathbf{k}+\mathbf{q}} | \frac{-e}{\sqrt{2\tilde{\omega}_{\mathbf{q}\mu}}} \sum_{l'\nu} \frac{\partial V_{l'}^c(\mathbf{r})}{\partial u_{\mathbf{q}l'\nu}} \frac{\partial u_{\mathbf{q}l'\nu}}{\partial U_{\mathbf{q}\mu}} | \psi_{l\mathbf{k}} \rangle$$

$$= \sum_{l'\nu} \frac{\tilde{e}_{\mathbf{q}\mu;l'\nu}}{\sqrt{2\tilde{\omega}_{\mathbf{q}\mu}}} \frac{ie^2}{A_{l'}} \frac{1}{\sqrt{N_{l'}}} |\mathbf{q}| z_{l'\nu} \langle \psi_{l\mathbf{k}+\mathbf{q}} | W_{l'}(\mathbf{q},z) e^{i\mathbf{q}\cdot\mathbf{r}} | \psi_{l\mathbf{k}} \rangle \tag{27}$$

In Eq. 27, we assume the |**q**| is small and thus we use $W_{l'l}(\mathbf{q})$ (see Note S2 for details on how to calculate this from the QEH model) to approximate $\langle \psi_{l\mathbf{k}+\mathbf{q}} | W_{l'}(\mathbf{q},z) e^{i\mathbf{q}\cdot\mathbf{r}} | \psi_{l\mathbf{k}} \rangle$. The $\bar{g}^{\mathrm{F}}$ can be evaluated as:

$$\bar{g}^{\mathrm{F}}_{l\mu}(\mathbf{k},\mathbf{q}) = \sum_{l'\nu} \frac{\tilde{e}_{\mathbf{q}\mu;l'\nu}}{\sqrt{2\tilde{\omega}_{\mathbf{q}\mu}}} \frac{ie^2}{A_{l'}} \frac{1}{\sqrt{N_{l'}}} |\mathbf{q}| W_{l'l}(\mathbf{q}) z_{l'\nu} \tag{28}$$

Note that the calculation of $\bar{g}^{\mathrm{F}}$ from Eq. 28 actually depends on the BvK boundary conditions: it has $N_{l'}$ in Eq. 28. In realistic calculations, we prefer to eliminate the arbitrary dependence on $N_l$. In the following, we show how to define a new $g^{\mathrm{F}}_{l\mu}$ that is independent of $N_l$. First, we note that the Fröhlich scattering rates $\tau^{\mathrm{F}}$ have to be well-defined observables and do not depend on $N_l$:

$$\frac{1}{\tau^{F}_{l\mathbf{k}}} = \sum_{\mathbf{k}'\in K_l} S(\mathbf{k},\mathbf{k}')$$

$$= \sum_{\mathbf{k}'\in K_l} \sum_{\mu} \frac{2\pi}{\hbar} |\bar{g}^{\mathrm{F}}_{l\mu}(\mathbf{k},\mathbf{k}'-\mathbf{k})|^2 \, w_{\mu}(\mathbf{k},\mathbf{k}') \tag{29}$$

where $S(\mathbf{k},\mathbf{k}')$ is the transition rate from the initial state $\mathbf{k}$ to the final state $\mathbf{k}'$, $K_l$ denotes the set of reciprocal lattice grid points, and $w(\mathbf{k},\mathbf{k}')$ is the transition weight, including contributions from electron and phonon occupancy:

$$w_{l\mu}(\mathbf{k},\mathbf{k}') = (f_{l\mathbf{k}'} + n_{\mathbf{k}'-\mathbf{k},\mu})\delta(E_{l\mathbf{k}'} - E_{l\mathbf{k}} - \hbar\tilde{\tilde{\omega}}_{\mathbf{k}'-\mathbf{k},\mu}) + (1 - f_{l\mathbf{k}'} + n_{\mathbf{k}'-\mathbf{k},\mu})\delta(E_{l\mathbf{k}'} - E_{l\mathbf{k}} + \hbar\omega_{\mathbf{k}'-\mathbf{k},\mu}) \tag{30}$$

Here, $f_{l\mathbf{k}}$ is the Fermi distribution for the electronic state $\mathbf{k}$ in layer $l$, $n_{\mathbf{q}\mu}$ is the Bose distribution for the phonon mode $\mu$ with wavevector $\mathbf{q}$, and $\tilde{\omega}_{\mathbf{q}\mu}$ is the eigenvalue of the new collective vibration obtained from Eq. 20. In Eq. 29, we replace the summation on final state $\mathbf{k}'$ by an integral over the Brillouin zone, which yields:

$$
\begin{aligned}
\frac{1}{\tau_{l\mathbf{k}}^{F}} &= N_l \int_l \frac{d\mathbf{k}'}{A_{l,\mathrm{BZ}}} \sum_{\mu} \frac{2\pi}{\hbar} |\, \bar{g}_{l\mu}^{\mathrm{F}}(\mathbf{k},\mathbf{k}'-\mathbf{k}) |^2 \, w_{l\mu}(\mathbf{k},\mathbf{k}') \\
&= \int_l \frac{d\mathbf{k}'}{A_{l,\mathrm{BZ}}} \sum_{\mu} \frac{2\pi}{\hbar} \Big| \sum_{l'v} \frac{\tilde{e}_{\mathbf{q}\mu;l'v}}{\sqrt{2\tilde{\omega}_{\mathbf{q}\mu}}} \frac{ie^2}{A_{l'}} \frac{\sqrt{N_l}}{\sqrt{N_{l'}}} |\mathbf{q}| W_{l'l}(\mathbf{q}) z_{l'v} \Big|^2 \, w_{l\mu}(\mathbf{k},\mathbf{k}') \\
&= \int_l \frac{d\mathbf{k}'}{A_{l,\mathrm{BZ}}} \sum_{\mu} \frac{2\pi}{\hbar} \Big| \sum_{l'v} \frac{\tilde{e}_{\mathbf{q}\mu;l'v}}{\sqrt{2\tilde{\omega}_{\mathbf{q}\mu}}} \frac{ie^2}{\sqrt{A_l A_{l'}}} |\mathbf{q}| W_{l'l}(\mathbf{q}) z_{l'v} \Big|^2 \, w_{l\mu}(\mathbf{k},\mathbf{k}')
\end{aligned} \tag{31}
$$

using the BvK boundary condition $A_l N_l = A_{l'} N_{l'}$. Eq. 31 can be written in a simpler form:

$$
\frac{1}{\tau_{l\mathbf{k}}^{F}} = \int_l \frac{d\mathbf{k}'}{A_{l,\mathrm{BZ}}} \sum_{\mu} \frac{2\pi}{\hbar} |\, g_{l\mu}^{\mathrm{F}}(\mathbf{k},\mathbf{q}) |^2 \, w_{l\mu}(\mathbf{k},\mathbf{k}') \tag{32}
$$

where $g_{l\mu}^{\mathrm{F}}(\mathbf{k},\mathbf{q})$ is defined as:

$$
g_{l\mu}^{\mathrm{F}}(\mathbf{k},\mathbf{q}) = \sum_{l'v} \frac{\tilde{e}_{\mathbf{q}\mu;l'v}}{\sqrt{2\tilde{\omega}_{\mathbf{q}\mu}}} \frac{ie^2}{\sqrt{A_l A_{l'}}} |\mathbf{q}| W_{l'l}(\mathbf{q}) z_{l'v} \tag{33}
$$

Eq. 33 is the computational equation for Fröhlich EPC matrix element (Eq. 5 in the main text) and the corresponding Fröhlich scattering rate can be obtained from Eq. 32. Another version more commonly used in literature and the equation for the corresponding mobility are provided in Note S3.

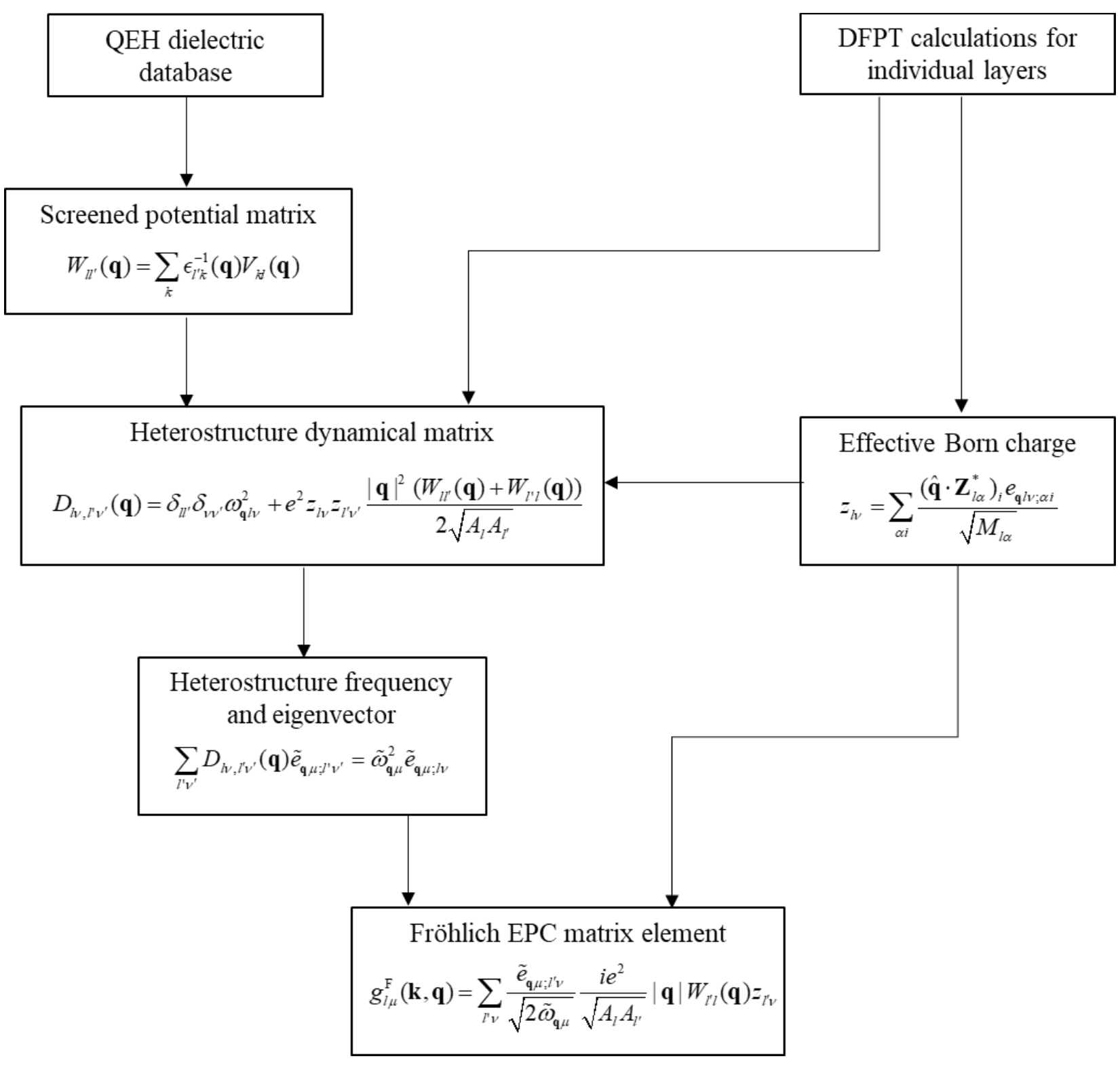

**Figure S1** Computation workflow of our model, from first-principles inputs: QEH dielectric database and DFPT calculations for individual layers, to Fröhlich EPC matrix element.

## Supplementary Note S2: Dielectric properties from the QEH model

In Note S1, we used the screened potential matrix $W_{ll'}(\mathbf{q})$ to accurately describe the dielectric screening in vdW heterostructures, both in the dynamical matrix (Eq. 19) and in the Fröhlich coupling matrix (Eq. 33). The screened potential matrix $W_{ll'}(\mathbf{q})$ quantifies the screened potential at layer $l'$ induced by bare charge at layer $l$:

$$W = \epsilon^{-1} V \tag{34}$$

where $\epsilon^{-1}$ is the inverse dielectric function and $V$ is the Coulomb kernel. In the QEH model, $\epsilon^{-1}$ and $V$ are calculated using a mixed density/potential basis [4]:

$$\begin{aligned} \epsilon^{-1}_{l\alpha,l'\alpha'} &= \langle \rho_{l\alpha} | \epsilon^{-1} | \phi_{l'\alpha'} \rangle \\ V_{l\alpha,l'\alpha'} &= \langle \rho_{l\alpha} | V | \rho_{l'\alpha'} \rangle \end{aligned} \tag{35}$$

where $l$, $l'$ are layer indices in the vdW heterostructure and $\alpha$, $\alpha'$ indices the monopole and dipole basis. (The definition of the density/potential basis can be found in Eqs. 13-15 and 18 in the Supporting Information of Ref. [4]) Here, we are interested in Fröhlich scattering, which only involves in-plane lattice vibrations and a "monopole"-like charge distribution. Therefore, we omit the $\alpha$ subscript shown in Eq. 35 in what follows. Using the same basis, the $W$ in Eq. 34 can be calculated via QEH model by:

$$W_{ll'}(\mathbf{q}) = \langle \rho_{l'} | W | \rho_l \rangle = \sum_k \epsilon^{-1}_{l'k}(\mathbf{q}) V_{kl}(\mathbf{q}) \tag{36}$$

where the $\epsilon^{-1}$ and $V$ matrices are calculated by the QEH code. The screened interaction $\overline{W}_{ll'}$ used in dynamical matrix is defined as:

$$\overline{W}_{ll'}(\mathbf{q}) = \frac{W_{ll'}(\mathbf{q}) + W_{l'l}(\mathbf{q})}{2} \tag{37}$$

When $l=l'$, the $W_{ll'}$ can be approximated by:

$$W_{ll}(\mathbf{q}) = \frac{2\pi}{|\mathbf{q}|(1 + 2\pi\alpha^{\parallel}|\mathbf{q}|)} \tag{38}$$

for isotropic 2D materials, where $\alpha^{\parallel}$ is the in-plane electronic polarizability [7]. Based on Eq. 19 or Ref. [7], the POP frequency for an individual layer $l$ with $W_{ll}(\mathbf{q})$ from Eq. 38 is given by:

$$\omega^2_{\mathbf{q}l\nu} = \omega^2_{0l\nu} + \frac{e^2 z^2_{l\nu}}{A_l} \frac{2\pi|\mathbf{q}|}{1 + 2\pi\alpha^{\parallel}|\mathbf{q}|} \tag{39}$$

Therefore, for relatively large $|\mathbf{q}|$, from Eq. 39 we have:

$$\omega^2_{\mathbf{q}\to\infty,l\nu} = \omega^2_{0l\nu} + \Delta\omega^2_{l\nu} \tag{40}$$

where $\Delta\omega$ is Eq. 6 in the main text:

$$\Delta\omega_{\text{POP}} = \frac{ez_{\text{POP}}}{\sqrt{\alpha^{\parallel} A}} \tag{41}$$

Eq. 39 shows that $\Delta\omega$ characterizes the frequency increase of POP near the zone center. In addition, Eq. 19 reveals that $\Delta\omega$ is proportional to the off-diagonal terms of the dynamical matrix of the vdW heterostructure, thereby quantifying the strength of interlayer POP coupling.

**Supplementary Note S3: Fröhlich and total mobility**

In Note S1, we demonstrated how to evaluate the Fröhlich EPC matrix element $g^{\text{F}}$ and the corresponding Fröhlich relaxation time $\tau^{\text{F}}$ in vdW heterostructures (see Eqs. 32 and 33). In this section, we supplement the equations for the total and Fröhlich mobility as part of the complete computational framework and provide additional calculation details. According to Matthiessen's rule, the total mobility $\mu$ can be evaluated in terms of the Fröhlich mobility $\mu^{\text{F}}$ and the non-Fröhlich mobility $\mu^{\text{NF}}$ as:

$$\frac{1}{\mu_l} = \frac{1}{\mu_l^{\text{F}}} + \frac{1}{\mu_l^{\text{NF}}} \tag{42}$$

Here, $\mu_l^{\text{F}}$ is the mobility limited by Fröhlich scattering in layer $l$, while $\mu_l^{\text{NF}}$ is determined by other scattering mechanisms. Surrounding materials can affect carrier mobility through various ways, including modifying the frequencies and EPC matrix elements for flexural modes [8, 9], or introducing additional short-range scattering such as surface roughness scattering [10, 11]. However, due to the absence of dangling bonds at the surfaces of 2D vdW materials, we assume that surrounding materials primarily influence electron transport via long-range Coulomb interactions. Therefore, in this work, only the scattering due to long-range interactions is considered to be modified. More specifically, since Fröhlich scattering induced by POPs constitutes the leading-order contribution to long-range scattering [12-14], we assume that the surrounding environment only modifies $\mu^{\text{F}}$, while $\mu^{\text{NF}}$ remains unchanged across different surroundings. This approximation significantly simplifies the computation: we only need to compute $\mu^{\text{F}}$ under various dielectric environments, while keeping $\mu^{\text{NF}}$ fixed. This assumption is particularly valid for polar semiconductors like InSe, whose mobility is primarily limited by Fröhlich scattering [12, 15]. The $\mu^{\text{F}}$ can be derived from Boltzmann transport theory within the relaxation time approximation [12, 16]:

$$\mu_{l,\alpha\beta}^{(\text{F})} = \frac{q}{n_c A_l} \int \frac{d\mathbf{k}}{A_{l,\text{BZ}}} \frac{\partial f_{l\mathbf{k}}}{\partial E_{l\mathbf{k}}} \tau_{l\mathbf{k}}^{(\text{F})} v_{l\mathbf{k},\alpha} v_{l\mathbf{k},\beta}, \tag{43}$$

where $\alpha$ and $\beta$ are the Cartesian directions, $q$ is the carrier charge, $A$ ($A_{\text{BZ}}$) is the area of the unit cell (Brillouin zone), and $n_c$ is the carrier density; $f_{n\mathbf{k}}$ is the Fermi distribution of the electronic state with band index $n$ and wavevector $\mathbf{k}$, $v_{n\mathbf{k}}$ is its group velocity, and $E_{n\mathbf{k}}$ is its electronic energy. $\tau^{\text{F}}$ is the Fröhlich relaxation time as given in Eq. 32. For consistency with previous works [12, 17], we rewrite Eq. 32 in terms of an integral over phonon wavevector $\mathbf{q}$ within the Brillouin zone:

$$\frac{1}{\tau_{\mathbf{k}}^{\mathrm{F}}}=\frac{2\pi}{\hbar}\sum_{\mu}\int_{\mathrm{BZ}}\frac{d\mathbf{q}}{A_{\mathrm{BZ}}}|g_{\mu}^{\mathrm{F}}(\mathbf{k},\mathbf{q})|^{2}[(f_{\mathbf{k+q}}+n_{\mathbf{q}\mu})\delta(E_{\mathbf{k}}-E_{\mathbf{k+q}}+\hbar\tilde{\omega}_{\mathbf{q}\mu}) \\ +(1+n_{\mathbf{q}\mu}-f_{\mathbf{k+q}})\delta(E_{\mathbf{k}}-E_{\mathbf{k+q}}-\hbar\tilde{\omega}_{\mathbf{q}\mu})] \tag{44}$$

where we neglect the layer index $l$ since we focus on transport within a single semiconductor layer of in the vdW heterostructure. The electronic band indices are also omitted here for simplicity. Although Eq. 44 formally resembles expressions used for suspended monolayers, the effects of surrounding materials are fully captured by the dependence of $g^{\mathrm{F}}$ and $\tilde{\omega}_{\mathbf{q}\mu}$ on the heterostructure dielectrics. In addition, we adopt the effective mass approximation to describe the electronic dispersion near the CBM of InSe, which is sufficient to capture the essential transport behavior in this regime.

**Supplementary Note S4: Model validation from first-principles calculations**

In this section, we use direct first-principles calculations to verify our model for heterostructures. A $ZrS_2/HfS_2$ bilayer heterostructure is constructed, and density functional perturbation theory (DFPT) calculations are performed to obtain the EPC $g$ matrix and phonon frequencies of the entire structure. $ZrS_2$ and $HfS_2$ are selected due to their similar lattice constants and large Born effective charges, which ensure strong POP coupling between the two layers. In practice, the first-principles calculations are performed using the Quantum Espresso Package [18, 19] with SG15 Optimized Norm-Conserving Vanderbilt (ONCV) pseudopotentials [20, 21] and the Perdew-Burke-Ernzerhof (PBE) exchange-correlation functional [22]. The 2D Coulomb cutoff technique [23] is applied to avoid fictitious EPC between image charges in DFPT calculations [24].

Figure S2 shows the calculated EPC strengths and phonon frequency for selected optical phonon modes. Figures S2 a and b compare the EPC strengths in $ZrS_2$ obtained from three approaches: first-principles DFPT, our model without interlayer POP coupling (using “DS + RPS” model; see main text for model details) and our complete model including interlayer POP coupling (using “DS + RPS + IPC” model; see main text for model details). Since $ZrS_2$ and $HfS_2$ both exhibit anisotropic effective mass, we compare EPC strengths for the same initial state at conduction band minimum (CBM) but with different final states along different directions, as illustrated in the insets. For both directions, the “DS + RPS + IPC” model (blue lines) agrees well with the DFPT benchmarks (red circles), especially for the two hybrid LO phonon modes (labeled ‘$LO_1$’ and ‘$LO_2$’), thereby validating the accuracy of our computational model. In contrast, the “DS + RPS” model (orange dashed lines), which omits interlayer POP coupling, exhibits significant deviations, especially for $LO_1$ near $\mathbf{q}=0$. This comparison directly demonstrates that IPC reduces Fröhlich scattering for one LO mode while enhancing it for the other. Figures S2 c and d show similar EPC trends in $HfS_2$. Finally, Figure S2 e and f compare the phonon frequencies of all optical modes from DFPT and from our model. For $q$-range considered near the zone center, all non-longitudinal optical modes exhibit smooth dispersion, which verifies the assumption we made in Eq. 19 (replacing $\omega(\mathbf{q})$ by $\omega(\mathbf{q}=0)$). In addition to accurately capturing the EPC strengths, our model also reproduces the $q$-dependence of the LO mode frequencies for both $\mathbf{q}$ directions.

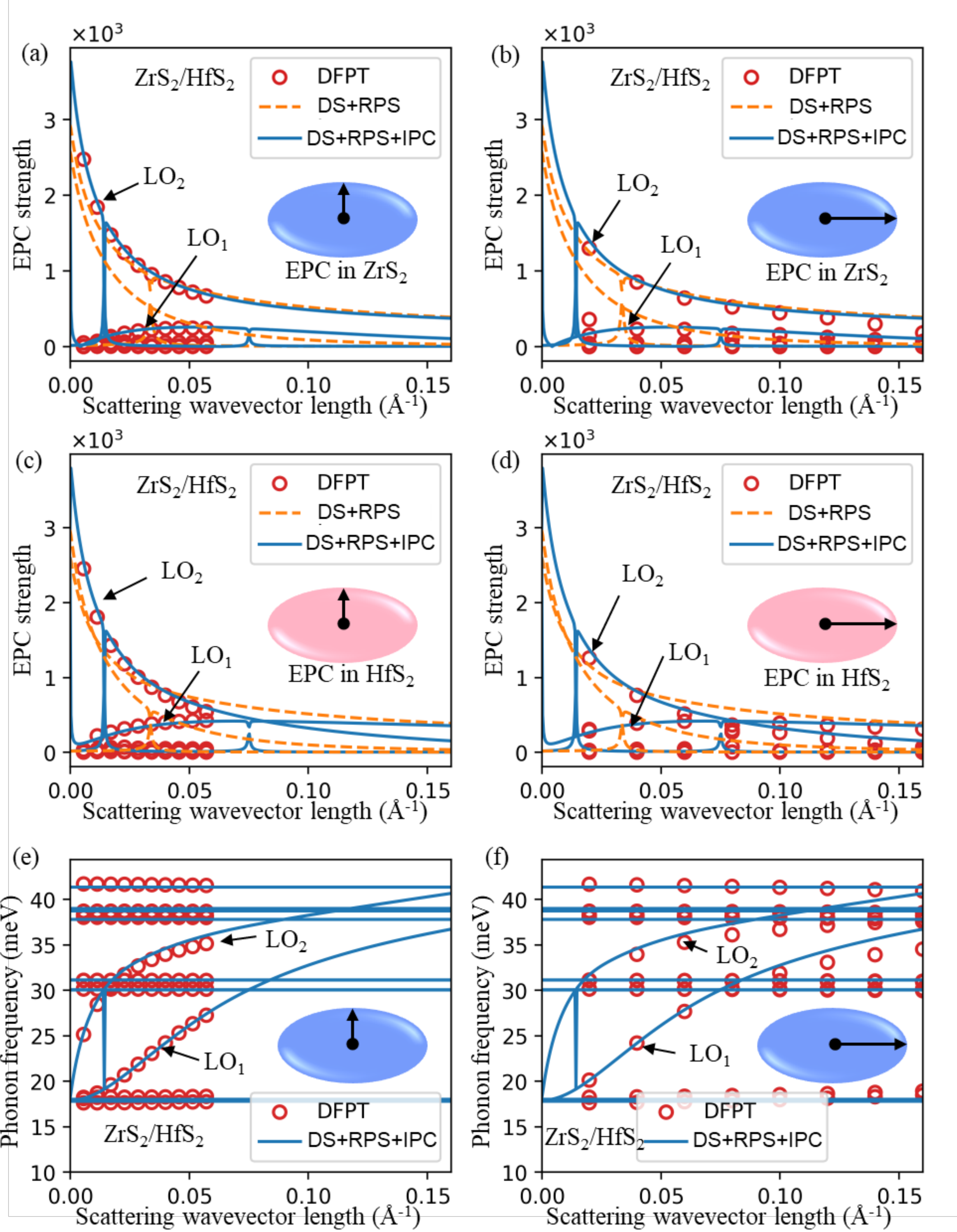

**Figure S2** (a-b) Electron-phonon coupling (EPC) strengths in $ZrS_2$ in the $ZrS_2/HfS_2$ bilayer heterostructure, calculated from first principles (DFPT), our model with interlayer POP coupling ("DS + RPS + IPC" ; see main text for model details), and without interlayer POP coupling ("DS + RPS"; see main text for model details) for phonon wavevector $\mathbf{q}$ along the $x$-direction (a) and $y$-direction (b). (c) and (d) show corresponding EPC strengths in $HfS_2$. (e) and (f) present the phonon dispersion of $ZrS_2/HfS_2$ bilayer hybrid optical modes for **q** along *x*- and *y*-directions, respectively.

## Supplementary Note S5: Frequency dispersion and Fröhlich scattering in h-BN/InSe vdW heterostructure

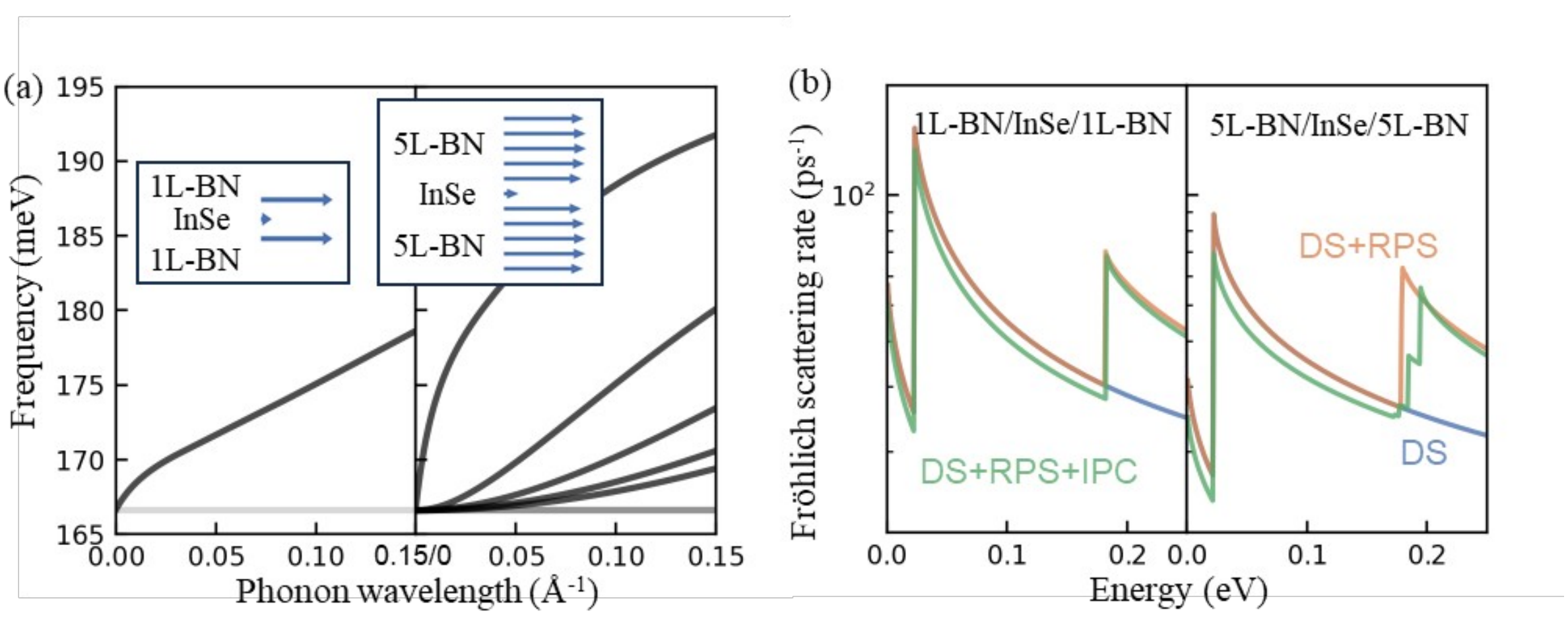

**Figure S3** (a) Frequency dispersion for hybrid POPs in 1L-BN/InSe/1L-BN and 5L-BN/InSe/5L-BN vdW heterostructures. Dispersionless non-POP frequencies are indicated by grey lines. Insets show the layer-resolved vibration magnitudes calculated from our model for the highest-frequency POP mode displayed. (b) Fröhlich scattering rates of InSe electrons in 1L-BN/InSe/1L-BN and 5L-BN/InSe/5L-BN vdW heterostructures, calculated using the "DS", "DS + RPS" and "DS + RPS + IPC" models (see main text for model details).

Figure S3 presents the frequency dispersion of hybrid POPs and the corresponding Fröhlich scattering for InSe in 1L-BN/InSe/1L-BN and 5L-BN/InSe/5L-BN vdW heterostructures. Compared to 1L-BN encapsulation, 5L-BN results in a more pronounced reduction in Fröhlich scattering for the first scattering rate peak at ~0.02 eV (corresponding to hybrid POP dominated by InSe; Figure S3b), and exhibits more complex POP vibration patterns within the heterostructure (Figure S3a). In the right panel of Figure S3b, the second scattering peak at ~0.17 eV (corresponding to h-BN-dominated POP scattering) is split into multiple smaller peaks, which is due to the splitting of degenerate POP frequency as shown in Figure S3a. The insets in Figure S3a illustrate the layer-resolved magnitude of POP vibrations calculated from our model, corresponding to the highest-frequency hybrid POP mode represented in the plot.

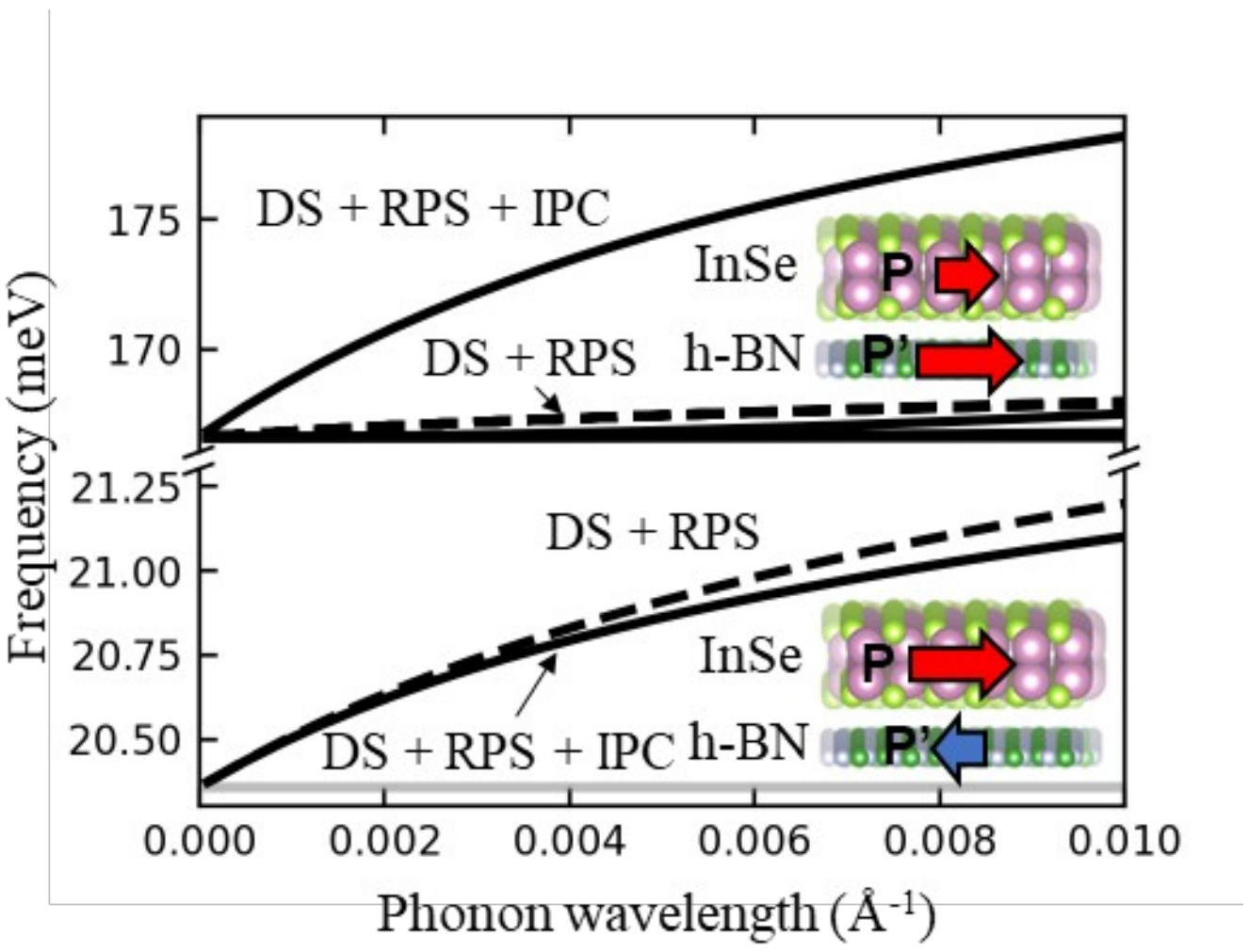

**Figure S4** Phonon dispersion of the 5L-BN/InSe/5L-BN vdW heterostructure, calculated with POP coupling ("DS + RPS + IPC" model; solid line) and without POP coupling ("DS + RPS" model; dashed line). Dispersionless non-POP frequencies are indicated by grey lines. Insets show the InSe layer and one h-BN layer in the heterostructure, with arrows indicating the electrostatic field generated by their POP vibrations.

## Supplementary Note S6: Complete list of 37 potential dielectric 2D layers

As discussed in the main text, we screened out 37 potential dielectric 2D layers from the Computational 2D Materials Database (C2DB) [25, 26]. However, only 5 of these materials are included in the QEH database, while the remaining 32 lack first-principles dielectric functions required for our model. Therefore, we use h-BN's dielectric properties to substitute that of 37 dielectric candidates, and evaluate the Fröhlich mobility ($\mu^{\mathrm{F}}$) of monolayer InSe encapsulated by these dielectrics (7L on each side). The calculated $\mu^{\mathrm{F}}$ from our model for these 37 dielectrics, along with their respective lowest zone-center POP frequency ($\omega_0$) and interlayer POP coupling strength ($\Delta\omega$; see main text for definition), are presented in Table S1 and Figure S5. In Figure S5, the grey dashed line indicates an iso-energy contour of 259 $\mathrm{cm^2V^{-1}s^{-1}}$, corresponding to the InSe $\mu^{\mathrm{F}}$ calculated using the dielectric screening (DS) model with h-BN encapsulation (7L on each side). Candidates located above this contour indicate that remote phonons in these dielectrics would improve the mobility of InSe.

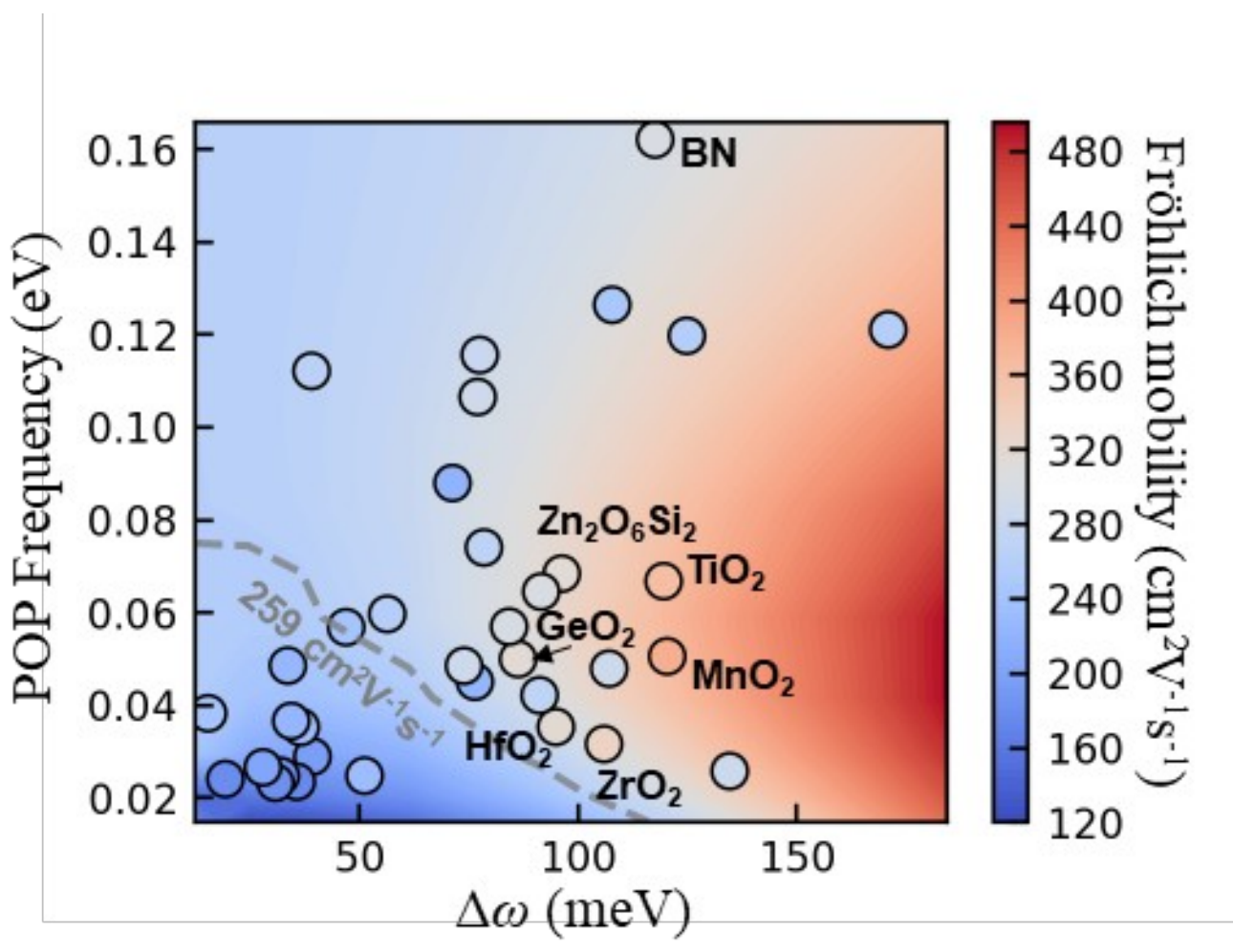

**Figure S5** InSe Fröhlich mobility ($\mu^{\mathrm{F}}$) with varied zone-center POP frequency $\omega_0$ and $\Delta\omega$ (see Eq. for definition) of surrounding 2D materials (7L on each side), for 37 promising 2D dielectric insulators. Dots represent individual dielectrics, colored according to InSe $\mu^{\mathrm{F}}$ in the corresponding heterostructures. Background color shows mobility calculated from a modified InSe/h-BN heterostructure model (7L on each side), with $\omega_0$ and $\Delta\omega$ of h-BN varied by scaling its dynamical matrix and Born effective charges, respectively. Grey dashed iso-mobility contour indicates the $\mu^{\mathrm{F}}$ of InSe encapsulated by h-BN, calculated using dielectric screening (DS) model.

| Formula | Space group | HSE bandgap (eV) | $\omega_0$ (meV) | $\Delta\omega$ (meV) | $\mu^{\mathrm{F}}$ ($cm^2V^{-1}s^{-1}$) |
|---|---|---|---|---|---|
| **BN*** | P-6m2 | 5.675 | 166.70 | 117.71 | 298.91 |
| **$ZrO_2$*** | P-3m1 | 6.508 | 31.48 | 87.75 | 315.78 |
| **$HfO_2$*** | P-3m1 | 6.592 | 35.07 | 85.09 | 335.74 |
| **$Ti_2O_4$** | P2_1/m | 5.343 | 24.26 | 82.19 | 232.33 |
| **AlN** | P-6m2 | 4.048 | 106.31 | 82.01 | 223.38 |
| **$GeO_2$*** | P-3m1 | 5.355 | 49.82 | 79.80 | 295.10 |
| **$MnO_2$** | P-3m1 | 4.134 | 50.41 | 73.32 | 280.27 |
| **$SnO_2$*** | P-3m1 | 4.204 | 48.14 | 65.31 | 261.60 |
| **$Zr_2O_6$** | Pmmn | 5.759 | 42.00 | 59.60 | 211.89 |
| **$Li_2H_2O_2$** | P4/nmm | 5.612 | 54.58 | 52.26 | 205.07 |
| **$Na_2H_2O_2$** | P1 | 5.180 | 35.09 | 45.87 | 124.32 |
| **$Ti_2O_6$** | Pmmn | 4.995 | 45.48 | 38.71 | 220.90 |
| **$CaCl_2$** | P-3m1 | 7.440 | 24.58 | 36.38 | 159.60 |
| **$MgCl_2$** | P-3m1 | 7.294 | 28.36 | 35.75 | 164.12 |
| **$GeO_2$** | P-4m2 | 4.862 | 28.13 | 35.66 | 181.41 |
| **$CaCl_2$** | P-6m2 | 6.384 | 23.12 | 33.45 | 156.21 |
| **$MgBr_2$** | P-3m1 | 5.744 | 23.33 | 33.16 | 151.40 |
| **$CH_2Si$** | P3m1 | 4.870 | 88.03 | 32.69 | 219.05 |
| **$Li_2H_2S_2$** | P1 | 5.027 | 35.75 | 28.34 | 185.86 |

| **$Zn_2Ge_2O_6$** | P-31m | 4.446 | 43.62 | 27.86 | 228.00 |
|---|---|---|---|---|---|
| **$Na_2H_2S_2$** | P1 | 4.587 | 23.90 | 25.36 | 155.96 |
| **$Y_2Cl_2O_2$** | P-3m1 | 6.000 | 20.01 | 22.47 | 186.20 |
| **$ZnH_2O_2$** | C2/m | 4.267 | 67.75 | 20.56 | 205.41 |
| **$Na_2H_2O_2$** | P2_1/m | 4.458 | 24.65 | 20.52 | 124.32 |
| **$NiCl_2$** | P-3m1 | 4.071 | 26.40 | 20.26 | 170.49 |
| **$Zn_2O_6Si_2$** | P-31m | 5.199 | 38.03 | 20.21 | 239.74 |
| **$HfO_2$** | P-4m2 | 6.108 | 24.87 | 17.70 | 212.71 |
| **$ZrO_2$** | P-4m2 | 6.238 | 25.15 | 15.68 | 231.44 |
| **$C_2F_2$** | P-3m1 | 4.908 | 37.48 | 15.61 | 195.87 |
| **$Au_2WO_4$** | P1 | 4.436 | 71.12 | 15.50 | 209.39 |
| **$MgH_2S_2$** | P1 | 4.551 | 35.87 | 14.37 | 206.10 |
| **$Li_2Cl_2O_4$** | P-42_1m | 5.196 | 56.69 | 14.35 | 195.97 |
| **$TiO_2$** | P-4m2 | 5.767 | 30.25 | 13.11 | 268.79 |
| **$Li_2H_4N_2O_6$** | Pc | 4.554 | 30.31 | 9.50 | 191.23 |
| **$Cr_2Cl_6$** | P-31m | 4.045 | 29.48 | 7.07 | 184.12 |
| **$ZrH_2O_6P_2$** | P-3m1 | 6.395 | 23.70 | 5.65 | 198.68 |
| **$C_2H_2$** | P-3m1 | 4.421 | 138.71 | 1.78 | 206.33 |

*Materials having the dielectric properties data in the QEH database

**Table S1** Complete list of 37 potential dielectric 2D layers, including their space group, HSE bandgap, $\omega_0$ and $\Delta\omega$. The reported $\mu^{\mathrm{F}}$ correspond to InSe Fröhlich mobility when monolayer InSe is encapsulated by these dielectric layers (7L on each side), calculated using the full model ("DS+RPS+IPC"). It should be noted that $\Delta\omega$ and $\mu^{\mathrm{F}}$ in this table are evaluated using the h-BN dielectric properties, due to the absence of first-principles dielectric functions in QEH database for most of the materials.

### Supplementary Note S7: Mobility degradation for encapsulated $MoS_2$

The mobility enhancement from remote phonons discussed in the main text focuses on the polar 2D semiconductor, InSe. As shown in Figure 4a in the main text, a large $\Delta\omega$ (defined in Eq. 6 in the main text) for the dielectric layer is critical for a strong interlayer POP coupling (IPC) and thus mobility enhancement. Similarly, a larger IPC also requires that the semiconductor layer itself has a sufficiently large $\Delta\omega$. Otherwise, a smaller $\Delta\omega$ implies less Fröhlich scattering to screen, and a reduced IPC effect, both of which suppress mobility enhancement. In this section, we substitute InSe ($\Delta\omega$ = 33 meV) with $MoS_2$ ($\Delta\omega$ = 14 meV) in dielectric encapsulations of h-BN, $SnO_2$, $GeO_2$, $ZrO_2$ and $HfO_2$, and evaluate the total mobility of $MoS_2$ in corresponding heterostructures. The $MoS_2$ total mobilities calculated from different models are shown in Figure S6. For $MoS_2$, total mobility is primarily limited by non-Fröhlich mobility ($\mu^{\mathrm{NF}}$), which we assume to remain constant across environments. Consequently, dielectric screening (DS) effect provides only minimal mobility enhancement across all dielectrics. In contrast, remote POP scattering (RPS) notably degrades the mobility, and the IPC is too weak (due to lower $\Delta\omega$ in $MoS_2$) to compensate for this mobility degradation. As a result, encapsulating $MoS_2$ with $SnO_2$, $GeO_2$, $ZrO_2$ or $HfO_2$ leads to mobility reductions ranging from 6% to 18%, in contrast to InSe where mobility is enhanced. Such discrepancy highlights the importance of considering semiconductor-dependent environmental effects when evaluating carrier mobility in vdW heterostructures.

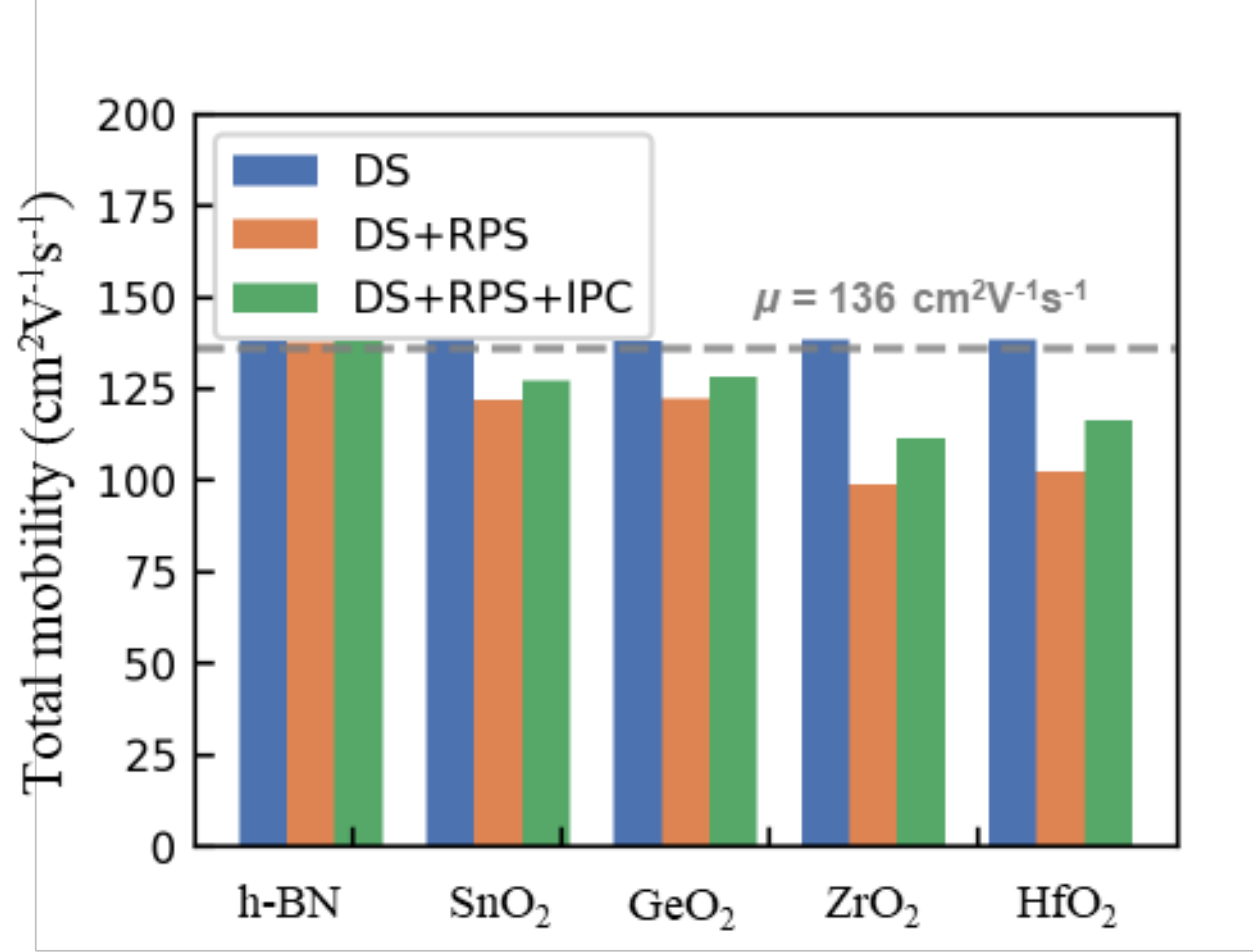

**Figure S6** Total mobility of $MoS_2$ encapsulated by $SnO_2$, $GeO_2$, $ZrO_2$ and $HfO_2$ (7L on each side), calculated using different models. The total mobility ($\mu$), Fröhlich mobility ($\mu^F$) and non-Fröhlich mobility ($\mu^{NF}$) for freestanding monolayer $MoS_2$ are 136 [12], 3950 and 141 $cm^2V^{-1}s^{-1}$, respectively.